\documentclass[pra,aps,twocolumn,showpacs,nofootinbib]{revtex4-1}
\usepackage{graphicx}
\usepackage{dcolumn}
\usepackage{bm}
\usepackage[bbgreekl]{mathbbol}
\usepackage{amsmath}

\def\ii{{\rm i}}  \def\ee{{\rm e}}  
      
\def\jb{{\bf j}}    \def\Eb{{\bf E}}  
    
\def\thb{\vec{\bf{\theta}}}

\begin{document}
\title{Graphene Plasmonics: Challenges and Opportunities}
\author{F.~Javier~Garc\'{\i}a~de~Abajo}
\email{javier.garciadeabajo@icfo.es}
\affiliation{ICFO-Institut de Ciencies Fotoniques, Mediterranean Technology Park, 08860 Castelldefels (Barcelona), Spain, and ICREA-Instituci\'o Catalana de Recerca i Estudis Avan\c{c}ats, Passeig Llu\'{\i}s Companys, 23, 08010 Barcelona, Spain}

\begin{abstract}
Graphene plasmons are rapidly emerging as a viable tool for fast electrical manipulation of light. The prospects for applications to electro-optical modulation, optical sensing, quantum plasmonics, light harvesting, spectral photometry, and tunable lighting at the nanoscale are further stimulated by the relatively low level of losses and high degree of spatial confinement that characterize these excitations compared with conventional plasmonic materials, alongside the large nonlinear response of graphene. We start with a general description of the plasmonic behavior of extended graphene, followed by analytical methods that lead to reasonably accurate estimates of both the plasmon energies and the strengths of coupling to external light in graphene nanostructures, including graphene ribbons. Although graphene plasmons have so far been observed at  mid-infrared and longer wavelengths, there are several possible strategies to extend them towards the visible and near infrared, including a reduction in the size of the graphene structures and an increase in the level of doping. Specifically, we discuss plasmons in narrow ribbons and molecular-size graphene structures. We further formulate prescriptions based on geometry to increase the level of electrostatic doping without causing electrical breakdown. Results are also presented for plasmons in highly-doped single-wall carbon nanotubes, which exhibit similar characteristics as narrow ribbons and show a relatively small dependence on the chirality of the tubes. We further discuss perfect light absorption by a single-atom carbon layer, which we illustrate by investigating arrays of ribbons using fully analytical expressions. Finally, we explore the possibility of exploiting optically pumped transient plasmons in graphene, whereby the optically heated graphene valence band can sustain collective plasmon oscillations similar to those of highly doped graphene, and well-defined during the picosecond time window over which the electron is at an elevated temperature. In brief, we discuss a number of exciting possibilities to extend graphene plasmons towards the visible and near-infrared spectral regions and towards the ultrafast time domain, thus configuring a vast range of possibilities for fundamental studies and technological applications.
\end{abstract}
\maketitle

\section{Introduction}

Plasmons --the collective oscillations of valence electrons in conducting materials-- possess a number of appealing properties for photonic technologies \cite{P08_2}, the most salient of which are (1) their small spatial extension compared with the light wavelength, which has been exploited to achieve improved imaging resolution \cite{LLX07}; (2) their strong interaction with light, which is evidenced by a centenary tradition of generating colors through plasmon-supporting metal nanoparticle suspensions \cite{H1976}; and (3) the huge optical enhancements produced by this strong interaction, which upon external illumination result in near-field intensities $>10^5$ times larger than the incident light intensity, as inferred from surface-enhanced Raman scattering (SERS)  measurements \cite{paper125}. Control over the spectral and spatial properties of these collective excitations has advanced at an impressive pace in recent years \cite{GPM08,HLC11}. Equally impressive are their applications to ultrasensitive detection down to the single-molecule level \cite{KWK97}, improved photovoltaics \cite{AP10}, nanoscale photometry \cite{KSN11}, cancer therapy \cite{LLH05}, and nonlinear optics \cite{HPR10}, among other feats.

Highly doped graphene has recently emerged as a powerful plasmonic material that combines the appealing properties noted above with the ability of being electrically tunable. In its undoped state, the atomically thin carbon layer presents $\sim2.3\%$ broadband absorption \cite{MSW08} mediated by excitation of electron-hole pairs. However, when electrically doped, an optical gap opens up, the energy of which is proportional to the applied bias voltage. Gaps nearing 2\,eV have been reported \cite{CPB11}. A plasmon band is then showing up in this gap at frequencies that are highly dependent on the doping level, and consequently, the optical response is fastly controllable through gated injection of charge carriers.

Radical variations in the optical absorption features associated with the excitation of graphene plasmons have been already demonstrated over a wide spectral range down to the mid-infrared \cite{FAB11,JGH11,SKK11,YLC12,YLL12,paper212,BJS13}. Additionally, these plasmons are highly confined to small regions compared with the wavelength, as directly observed through scanning near-field optical microscopy \cite{paper196,FRA12}. It should be mentioned that graphene also exhibits higher-energy plasmons \cite{EBN08,WHK13} ($>5\,$eV) of limited tunability,  similar to those in metals, which we will not discuss here. Low-energy, tunable plasmons have been observed in extended graphene \cite{LWE08,LW10,KSS10,TPL11,SKK11,FAB11}, graphene ribbons \cite{JGH11,paper196,FRA12,YLZ13,SNC13}, disks \cite{YLL12,paper212,paper230}, rings \cite{YXL12,paper212}, disk stacks \cite{YXL12,YLC12}, and holes \cite{BJS13}; these observations have been carried out at IR \cite{LWE08,LW10,KSS10,TPL11,SKK11,FAB11,paper196,FRA12,paper212,BJS13,YLZ13,paper230} and THz \cite{JGH11,YLL12,YXL12,YLC12,SNC13} frequencies, using low-energy electron energy-loss \cite{LWE08,LW10,KSS10,TPL11,SKK11}, optical far-field \cite{JGH11,YLL12,YXL12,YLC12,paper212,BJS13,YLZ13,SNC13,paper230}, and scanning optical near-field \cite{FAB11,paper196,FRA12} spectroscopies. Strong changes in the graphene plasmon dispersion due to hybridization with the optical phonons of a SiC substrate have also been observed \cite{LW10,YLZ13}. Besides, graphene magnetoplasmons have been measured at THz frequencies \cite{COP12,YLL12}, similar to those of conventional two-dimensional electron gases \cite{F1986}, which can be tuned through varying the intensity of an externally applied magnetic field. These findinds have stimulated a tremendous activity purposing to explore and exploit the photonic and plasmonic properties of graphene, as discussed in recent reviews \cite{GPN12,BL12,BFP13}.

Experimental efforts in graphene plasmonics have been matched by an extensive wealth of theoretical analyses, including microscopic quantum descriptions based upon the random-phase-approximation (RPA) for extended graphene \cite{WSS06,HD07}, narrow ribbons \cite{BF07,paper183}, and other structures \cite{paper183,paper214}. Additionally, classical electrodynamic solutions have been produced for more complex geometries, such as graphene circuits \cite{VE11}, individual disks and ribbons \cite{paper176,NGG11,WK13}, dimers \cite{paper181,paper216}, periodically patterned layers \cite{paper182,BPV12,FP12,NGG12,paper211}, and tips \cite{RFM13}.

The unique plasmonic behavior of graphene, combined with its excellent electronic properties \cite{CGP09,BSH10}, has triggered a race to understand the dynamics of hot electrons in this material \cite{BTM13}, as well as the mechanisms leading to inelastic plasmon attenuation \cite{LHJ08,PRN10,PVC13}. Besides these fundamental aspects, graphene is attracting considerable interest because of its potential application to optical signal processing \cite{VE11}, light modulation \cite{LYU11}, sensing \cite{SGM07}, spectral photometry \cite{FLW12,VVC12}, quantum optics \cite{paper176,paper184}, and nonlinear photonics \cite{M07_2,HHM10,M11,WZY11}. This excitement originates in part in the large electro-optical response of the atomically thin carbon layer \cite{FAB11,JGH11,SKK11,YLC12,YLL12,paper212,BJS13,paper196,FRA12}. In contrast, conventional plasmonic materials are traditionally tuned through geometry \cite{GPM08}, as only massive amounts of chemically induced doping can produce observable plasmon shifts \cite{MPG06}, although some promising strategies are being explored to achieve plasmon tunability in nanoparticles \cite{CST14}.

Unfortunately, graphene plasmons have only been observed at mid-IR and longer wavelengths. Consequently, intense efforts are currently underway to exploit the graphene tunability at shorter wavelengths. In this context, exciting results have been reported on the electro-optical control of extrinsic plasmons sustained by noble metal nanostructures \cite{ECN12,YKG13,LY13,ECK14}, as well as light propagation in integrated silicon waveguides \cite{LYU11} and photonic crystal cavities \cite{GSG13}, on which an electrically pumped nearby graphene layer can drive observable spectral changes. Alternative strategies here discussed consist in raising the level of doping and reducing the size of the graphene structures. We further explore optically pumped transient plasmons and excitations in graphene-like molecular structures. Graphene is also argued to hold great potential both for obtaining exotic optical behavior, such as complete absorption within an atomically thin layer \cite{paper182} and for implementing quantum-optics devices in a solid-state environment \cite{paper186}.

\section{Fundamentals of Graphene Plasmonics}

\subsection{Optical Response and Plasmons in Graphene}

\begin{figure*}
\begin{center}
\includegraphics[width=170mm,angle=0,clip]{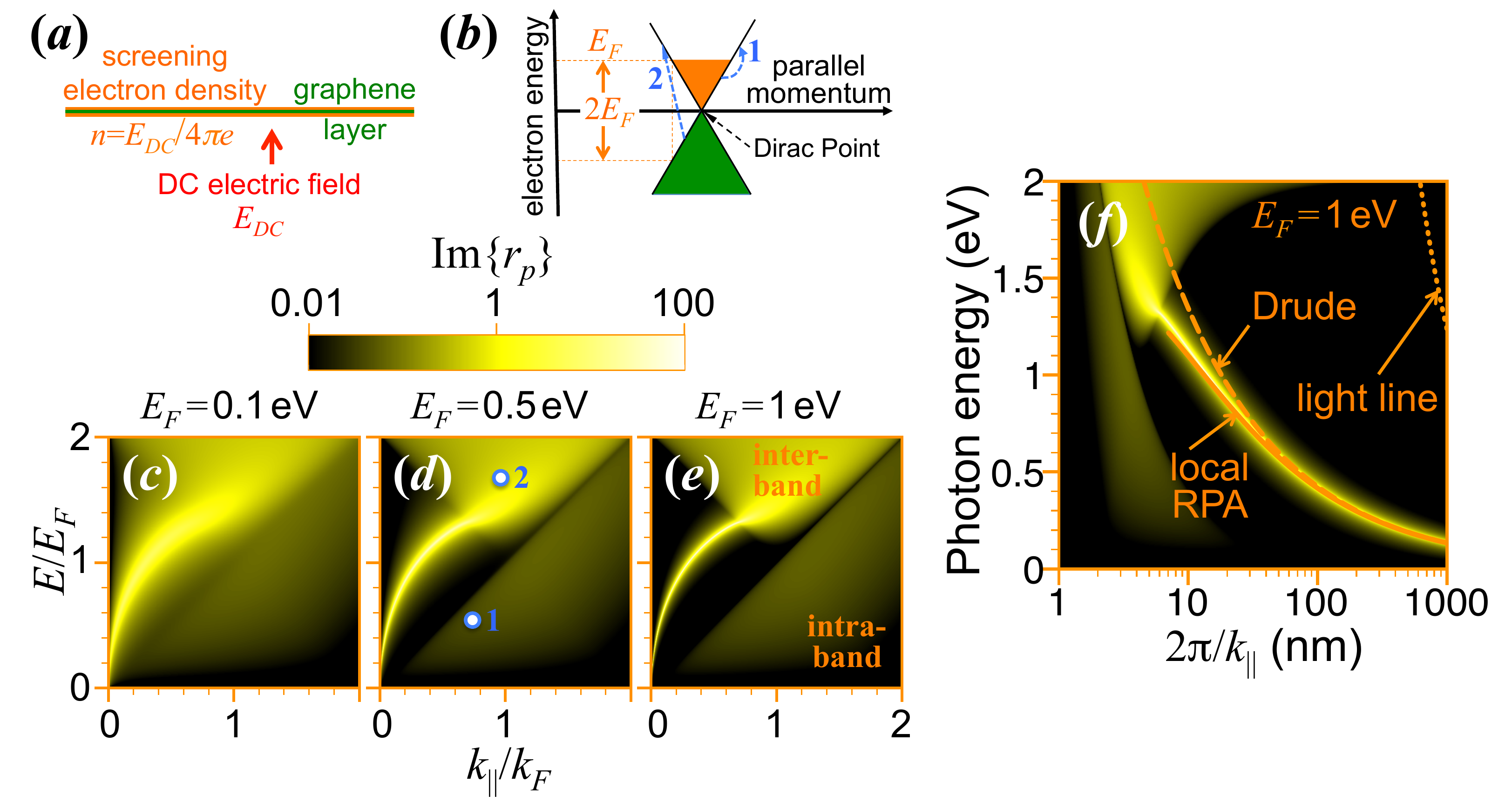}
\caption{{\bf Graphene electrostatic doping and plasmon dispersion relation.} {\bf (a)} An applied DC electric field, which is for example produced via backgating, induces {\it doping} charges on the graphene. {\bf (b)} A doping charge density $n$ raises the Fermi level to $E_F=\hbar v_F k_F$, with $k_F=\sqrt{\pi n}$, and consequently, a gap is opened of size $2E_F$ for vertical transitions. {\bf (c-e)} This optical gap closes down for parallel wave vector transfers $k_\parallel\ge k_F$ and a plasmon mode is allowed to exist free from Landau damping in the remaining $k_\parallel<k_F$ region. Here we visualize the gap and the plasmon by representing the $k_\parallel-\omega$ dependence of the loss function ${\rm Im}\{r_p\}$ for three different levels of doping in extended free-standing graphene. Two specific intraband (1) and interband (2) transitions are shown in (d), corresponding to the dashed arrows in (b). {\bf (f)} Same as (e), but represented as a function of in-plane wavelength $2\pi/k_\parallel$ rather than parallel wave vector $k_\parallel$. The light line (dotted curve) and the plasmon dispersion relations in the Drude (dashed curve) and local-RPA (solid curve) models are shown for comparison. The density plots are obtained using the full RPA conductivity for graphene \cite{WSS06,HD07} with mobility $\mu=2,000\,$cm$^2/$(V\,s).} \label{Fig1}
\end{center}
\end{figure*}

Near its Fermi level, the electronic band structure of graphene is dominated by two inequivalent singular points in the first Brillouin zone, each of them consisting of two cones with their tips touching right at the so-called Dirac points, which coincide with the Fermi energy of the undoped carbon layer \cite{W1947,CGP09}. Choosing the origin of 2D electron momentum $k_\parallel$ at one of the two Dirac points, the electron energy follows a linear dispersion $E\approx\hbar v_F k_\parallel$, where $v_F\approx10^6\,$m$/$s is the Fermi velocity. Injection of charge carriers (electrons or holes), through for example electrical gating \cite{CPB11} or chemical doping \cite{LLZ11}, moves the Fermi energy to $E_F=\hbar v_F k_F$, where $k_F=\sqrt{\pi n}$ is the Fermi wave vector and $n$ is the concentration of additional carriers (see Fig.\ \ref{Fig1}a,b). Under realistic conditions, electrical gating can be used to produce $E_F\sim1\,$eV \cite{CPB11}, which corresponds to $n\sim7\times10^{13}\,$cm$^{-2}$.

An immediate consequence of doping in graphene is the opening of an optical gap of size $2E_F$ for vertical transitions (see Fig.\ \ref{Fig1}c-f). We illustrate the opening of this gap by plotting in Fig.\ \ref{Fig1} a representative {loss function} for various values of $E_F$. In particular, we show ${\rm Im}\{r_p\}$, where $r_p$ is the Fresnel reflection coefficient for $p$-polarized light. This function accounts for the enegy-loss probability when the graphene is excited by a fast electron \cite{paper228}, but it also illustrates the wave vector $k_\parallel$ and frequency $\omega$ dependence of the optical excitations in this material, including the emergence of a plasmon band. It is useful to realize that retardation effects are negligible over the region of interest ({\it i.e.}, $k_\parallel<k_F$ and $\hbar\omega<2E_F$). Indeed, the Fermi wavelength is $\lambda_F=2\pi/k_F\sim4-10\,$nm for typical values of $E_F\sim0.4-1\,$eV, so that the light wavelength ($\lambda_0>1-3\,\mu$m for $\hbar\omega<E_F$) is much larger than $\lambda_F$ and the light line cannot be distinguished from the vertical $k_\parallel=0$ axis on the scale of Fig.\ \ref{Fig1}c-e. In the electrostatic limit, we find
\begin{equation}
r_p=\frac{1}{1-\ii\omega/(2\pi k_\parallel\sigma)}
\label{rp}
\end{equation}
in terms of the graphene conductivity $\sigma(k_\parallel,\omega)$. The density plots of Fig.\ \ref{Fig1} are obtained using the RPA for $\sigma(k_\parallel,\omega)$ \cite{W1947,WSS06,HD07}, which accounts for nonlocal effects within linear response theory \cite{PN1966}, using a tight-binding description for the $\pi$-band electron wave functions. As noted above, the gap narrows down under oblique light incidence ({\it i.e.}, for $k_\parallel\neq0$) and it completely disappears at $k_\parallel=k_F$. It is precisely in this triangular gap region where plasmons show up as a distinct absorption feature.

A representation in terms of the in-plane wavelength $2\pi/k_\parallel$ (Fig.\ \ref{Fig1}f) corroborates that the plasmons are far apart from the propagating light modes in the spectral range of interest. Consequently, we can generally neglect retardation effects in the analysis of graphene plasmons under high doping conditions. Actually, those effects cannot be resolved on the scale of Fig.\ \ref{Fig1}, thus justifying the use of the above electrostatic limit for Eq.\ (\ref{rp}). Unfortunately, this also implies that the coupling of graphene plasmons to propagating light becomes a challenge, which can however be overcome by patterning the graphene layer to boost light absorption, as discussed below.

One expects the $k_\parallel$ dependence to play a negligible role for graphene islands of size $\gg\lambda_F$. This intuition has been recently confirmed by full RPA calculations for finite graphene structures \cite{paper183}, which allow us to conclude that we can realistically model them within the local limit ($k_\parallel=0$). The RPA conductivity of extended graphene then reduces to the local-RPA conductivity
\begin{widetext}
\begin{equation}
\sigma(\omega)\equiv\sigma(k_\parallel=0,\omega)=\frac{-e^2}{\pi\hbar^2}\frac{\ii}{\omega+\ii\tau^{-1}}\int_{-\infty}^\infty dE\;\left[|E|\frac{\partial f_E}{\partial E}+\frac{(E/|E|)}{1-4E^2/[\hbar^2(\omega+\ii\tau^{-1})^2]}\;f_E\right],
\label{local}
\end{equation}
\end{widetext}
where $f_E$ is the electron distribution as a function of energy $E$, and we have used the prescription of Mermin \cite{M1970} to locally conserve electron density when a finite electronic relaxation time $\tau$ is introduced. The first term inside the integral of Eq.\ (\ref{local}) gives the contribution from intraband transitions ({\it i.e.}, electron transitions within the same partly filled Dirac cone, see labels (1) in Fig.\ \ref{Fig1}b,d), which at zero temperature ({\it i.e.}, for $f_E=\theta(E_F-E)$) produces the Drude conductivity
\begin{equation}
\sigma(\omega)=\frac{e^2}{\pi\hbar^2}\frac{\ii E_F}{\omega+\ii\tau^{-1}}.
\label{Drude}
\end{equation}
The second term in Eq.\ (\ref{Drude}) describes interband transitions across the optical gap (see labels (2) in Fig.\ \ref{Fig1}b,d). For plasmon energies $E_p<E_F$ and large structures compared with $\lambda_F$, this term can be neglected and Eq.\ (\ref{Drude}) yields a fairly good approximation. Additionlly, the intraband term admits an easily computable expression in the $\tau\rightarrow\infty$ limit \cite{FV07,FP07_2}, which provides a reasonably accurate correction due to interband polarization effects, and has been extensively used in the analysis of graphene plasmons \cite{paper176,paper212}. Throughout this paper, we retain instead the full finite $\tau$ dependence in Eq.\ (\ref{local}), as this becomes relevant under the extreme doping conditions here discussed, including ultrafast optical pumping leading to transient plasmons.

The plasmon dispersion relation of extended graphene is given by the poles of $r_p$ ({\it i.e.}, $k_{\rm sp}=\ii\omega/(2\pi\sigma)$). Using Eq.\ (\ref{Drude}) in this expression, we find the plasmon wavelength $\lambda_{\rm sp}=2\pi/{\rm Re}\{k_{\rm sp}\}$ to be related to the light wavelength $\lambda_0$ as
\begin{equation}
\frac{\lambda_{\rm sp}}{\lambda_0}=\frac{4\alpha}{\epsilon_1+\epsilon_2}\;\frac{E_F}{\hbar\omega},
\label{lamsp}
\end{equation}
where $\alpha=e^2/\hbar c\approx1/137$ is the fine-structure constant. For the sake of completeness, we have corrected this expression by adding a factor $2/(\epsilon_1+\epsilon_2)$ to the right-hand side in order to account for the effect of dielectric environment when the graphene is sitting at the planar interface between two dielectrics of permittivities $\epsilon_1$ and $\epsilon_2$ (see more details below). The dashed curve in Fig.\ \ref{Fig1}f is obtained from this expression with $\epsilon_1=\epsilon_2=1$ (self-standing graphene). The agreement with the RPA plasmon dispersion relation is excellent at low values of $k_\parallel$, again confirming the validity of the local approximation for large structures. Furthermore, electron-hole-pair excitations can be neglected in the dynamical interaction between induced charges at distances larger than a few tens of nanometers, as the plasmon dispersion relation only enters that region for shorter in-plane wavelengths $2\pi/k_\parallel$ (see Fig.\ \ref{Fig1}f).  Consequently, electron-hole-pair excitations should be also negligible in structures of sizes larger than those distances, in agreement with RPA calculations for finite graphene islands \cite{paper183}. Incidentally, polarization of interband transitions produces a plasmon shift, which is well described by the local-RPA conductivity (Fig.\ \ref{Fig1}f, solid curve).

\subsection{Optical Losses}

In the above expressions, we have introduced an intrinsic decay rate $\tau$ that causes the plasmons to acquire a finite lifetime and is influenced by several factors, such as collisions with impurities 
\cite{PVC13}, coupling to optical phonons \cite{JBS09}, and finite-size and edge effects \cite{paper183}. Each of these mechanisms provides additional momentum needed to break the mismatch between plasmons and electron-hole-pair excitations within the gaps of Fig.\ \ref{Fig1}. In particular, the DC Drude model \cite{AM1976} permits estimating the impurity-limited lifetime as $\tau=\mu E_F/ev_F^2$, where $\mu$ is the mobility. For reference, this expression predicts $\tau\approx1\,$ps ({\it i.e.}, $\hbar\tau^{-1}\approx0.66\,$meV) for $E_F=1\,$eV and $\mu=10,000\,$cm$^2/$(V\,s). Although even higher mobilities have been measured in both suspended \cite{NGM04,NGM05} and BN-supported \cite{DYM10} graphene, experimental plasmon studies \cite{FAB11,JGH11,SKK11,YLC12,YLL12,paper212,BJS13,paper196,FRA12} have so far reported lower $\mu$'s ($<2,000$), thus demanding cleaner fabrication methods for graphene patterning and device fabrication in order to meet the expectation of long-lived optical modes in defect-free graphene structures. Additionally, a proper treatment of impurity scattering beyond the Drude model \cite{PVC13} seems to explain the presence of a residual plateau in the measured losses within the optical gap \cite{LHJ08}. Moreover, intrinsic optical phonon losses have been predicted to dramatically reduce the plasmon lifetime for energies $E_p>0.2\,$eV \cite{JBS09}, in agreement with a recent study of graphene ribbons \cite{YLZ13}, but in contrast to the observation of narrow plasmons at energies above $0.3\,$eV in nanorings \cite{paper212}. Coupling to substrate phonons is another potential source of losses. Zigzag edges have also been found to produce dramatic plasmon damping due to the presence of electronic edge states \cite{paper183}. Quite differently, armchair nanoislands do not host such edge states, and therefore, their plasmons are expected to be narrower and better defined than in zigzag islands \cite{paper214}. For sufficiently small armchair islands down to molecular sizes, these plasmons can be even pushed to the visible and near-infrared (vis-NIR) \cite{paper215}.

\subsection{Electrostatic Scaling Laws}

Because $\lambda_{\rm sp}\ll\lambda_0$ [see Eq.\ (\ref{lamsp})], we can safely neglect retardation and express the optical response of graphene in terms of an electrostatic potential $\phi$. We consider a homogeneously doped graphene structure of characteristic size $D$ placed at the planar interface between two media of permittivities $\epsilon_1$ and $\epsilon_2$. Although the formalism presented below applies to any choice of $D$, it is convenient to associate it with a characteristic distance, such as the diameter in a disk, the side in a square, or the width in a ribbon. For finite arbitrary shapes, one could set $D$ to the square root of the graphene area. It is then convenient to define a filling function $f$ that takes the value 1 on the graphene and vanishes elsewhere. Using dimensionless coordinates $\thb=(x/D,y/D)$ on the graphene plane, we can write the self-consistent relation
\begin{equation}
\phi(\thb)=\phi^{\rm ext}(\thb)+\eta \int \frac{d^2\thb'}{|\thb-\thb'|}\,\nabla'\cdot f(\thb')\nabla'\phi(\thb'),
\label{phi}
\end{equation}
where $\phi^{\rm ext}$ is the external potential, whereas
\begin{equation}
\eta=\frac{i\sigma(\omega)}{\omega D} \frac{2}{\epsilon_1+\epsilon_2} \label{eta}
\end{equation}
is a dimensionless parameter. The integral term in Eq.\ (\ref{phi}) is just the Coulomb potential produced by the induced charge $\rho$, which is in turn expressed in terms of the induced current $\jb=-\sigma f\nabla\phi$ through the continuity equation $\rho=(-\ii/\omega)\nabla\cdot\jb$. Notice that the bare Coulomb potential $1/r$ due to a point charge must be corrected by a factor $2/(\epsilon_1+\epsilon_2)$ when the charge is sitting at the interface between two dielectrics. This factor is simply pulled out in front of the integral in Eq.\ (\ref{eta}) as an exact correction to account for the interface. Incidentally, Eq.\ (\ref{phi}) also describes inhomogeneously doped graphene in the Drude approximation \cite{paper194}, with the spacial dependence of the Fermi energy transferred into $f$.

The dependence on frequency, doping level, dielectric environment, and absolute size of the structure is fully contained in $\eta$. The rest of the elements in Eq.\ (\ref{phi}) have a purely geometrical interpretation. Following a similar approach as for general electrostatic problems \cite{OI1989}, this equation can be recast into a real-symmetric eigensystem \cite{paper228}, in which the plasmon resonances are identified with negative real eigenvalues $\eta_j$. The complex plasmon frequencies are then obtained by solving the equation $\eta=\eta_j$. In particular, using the Drude model [Eq.\ (\ref{Drude})], we find $\omega\approx\omega_j-\ii\tau^{-1}/2$, with
\begin{equation}
\hbar\omega_j=e\;\sqrt{\left(\frac{1}{-\pi\eta_j}\right)\left(\frac{2}{\epsilon_1+\epsilon_2}\right)}\;\sqrt{\frac{E_F}{D}}.
\label{wj}
\end{equation}
Given any geometrical shape, this scaling law allows us to obtain the plasmon energy for all desired values of $E_F$ and $D$, provided we know the energy for a specific choice of these parameters. For example, taking $E_F=1\,$eV  and $D=100\,$nm, the lowest-order dipole plasmon energy $\hbar\omega_j$ is $0.25\,$eV for a disk of diameter $D$ 
and $0.26\,$eV for a ribbon of width $D$ (dipole mode across the ribbon, see below).

From the above analysis we can obtain a scaling law for the polarizability $\alpha_\omega$ of a graphene island by considering an external potential $-E_0x$ corresponding to an external field $E_0$ along a symmetry direction $x$. We find \cite{paper212}
\begin{equation}
\alpha_\omega=D^3\sum_j\frac{A_j}{\frac{(-2/\eta_j)}{\epsilon_1+\epsilon_2}-\frac{\ii\omega D}{\sigma(\omega)}},
\label{alpha}
\end{equation}
where $A_j$ are real, positive, dimensionless coupling coefficients, which depend on the specific geometry under consideration (see Ref.\ \cite{paper212} for more details) and can be calculated once and for all by solving Eq.\ (\ref{phi}) and comparing the resulting polarizability with Eq.\ (\ref{alpha}). In the $\sigma\rightarrow0$ limit (weak coupling regime), substituting $\phi^{\rm ext}$ for $\phi$ in the integral of Eq.\ (\ref{phi}), we obtain the sum rule
\begin{equation}
\sum_jA_j=\frac{A}{D^2},
\label{sumAj}
\end{equation}
where $A$ is the graphene area. In the opposite limit ($|\sigma|\rightarrow\infty$), the graphene behaves as a perfect conductor of polarizability $\alpha_0$, which allows us to obtain a second sum rule,
\begin{equation}
-\sum_j\eta_jA_j=\frac{2(\alpha_0/D^3)}{\epsilon_1+\epsilon_2}.
\label{rule2}
\end{equation}
In particular, $\alpha_0/D^3=1/6\pi$ for a self-standing circular disk \cite{J99}. 
Incidentally, assuming that the lowest-energy plasmon dominates the above sums, we find $\eta_j\approx-2/3\pi^2$, which allows us to analytically predict a plasmon energy of $0.26\,$eV for the disk with $E_F=1\,$eV and $D=100\,$nm, in reasonable agreement with the numerical value of $0.24\,$eV quoted above.

Notice that Eq.\ (\ref{alpha}) can be readily used to obtain the absorption cross-section of a self-standing structure as $\sigma^{\rm abs}=(8\pi^2/\lambda_0){\rm Im}\{\alpha_\omega\}$. Incidentally, for undoped graphene, the conductivity reduces to $\sigma=e^2/4\hbar$ over a wide spectral range, leading to
\begin{equation}
\frac{\sigma^{\rm abs}}{A}=\pi\alpha\;\;g\left[\frac{8\pi D}{\alpha\lambda_0}\right],
\label{undoped}
\end{equation}
where $g(x)=x^2(D^2/A)\sum_jA_j/(x^2+1/\eta_j^2)$ is obviously a monotonically increasing function of $x\approx0.183\,D/\lambda_0$ (see Fig.\ \ref{Fig8}c). In the small structure limit, we have $g(0)=0$ (depletion of absorption), whereas $g\approx1$ for $D\gg\lambda_0$ in virtue of Eq.\ (\ref{sumAj}), thus resulting in an absorbance $\sigma^{\rm abs}/A\approx\pi\alpha$, in agreement with optical measurements of large graphene islands \cite{MSW08}.

To summarize, the above scaling laws emanate from Eq.\ (\ref{phi}), which provides a complete classical electrostatic description of graphene in the local approximation, with the size of the system $D$, the conductivity $\sigma$, and the frequency $\omega$ fully embedded inside the parameter $\eta$ [Eq.\ (\ref{eta})]. These laws are expressed in terms of the dimensionless constants $\eta_j$ and $A_j$, which are independent of size, conductivity, and frequency. Additionally, the plasmonic response is dominated by a single mode $j$ in many geometries, such as ribbons (see below), which allows us to derive rather accurate analytical expressions for the absorption cross-section, thus emphasizing the power of the scaling laws.

It shoud be noted that the above analysis relies on a description of the graphene as an infinitely thin layer of finite 2D conductivity $\sigma$. In practice, classical numerical simulations have been reported for several geometries by describing the graphene as a film of finite thickness $t$ \cite{VE11,paper176,paper181} and permittivity $\epsilon=1+4\pi\ii\sigma/(t\,\omega)$, so that the problem reduces to solving Maxwell's equations, using for example the boundary-element method (BEM) \cite{paper040,paper176}. Converged results in the $t\rightarrow0$ limit are obtained with $t\sim0.3\,$nm (the interlayer separation of graphite) for islands above 100\,nm in size. However, a small dependence on $t$ is still observable in smaller islands within this range of thicknesses. Nonetheless, the plasmon energies only differ by a small percentage from the $t=0$ values in islands as small as 10\,nm, for which finite-size quantum effects require moving to non-classical methods anyway, such as the TB+RPA (see Appendix\ \ref{Ap1}). Our numerical estimates of $\eta_j$ and $A_j$ are obtained using the above classical description. Incidentally, we also use an intrinsically $t=0$ alternative procedure based upon surface dipole elements to solve the electrostatic problem for ribbons in next paragraph (see Appendix\ \ref{Ap2}).

\begin{figure*}
\begin{center}
\includegraphics[width=150mm,angle=0,clip]{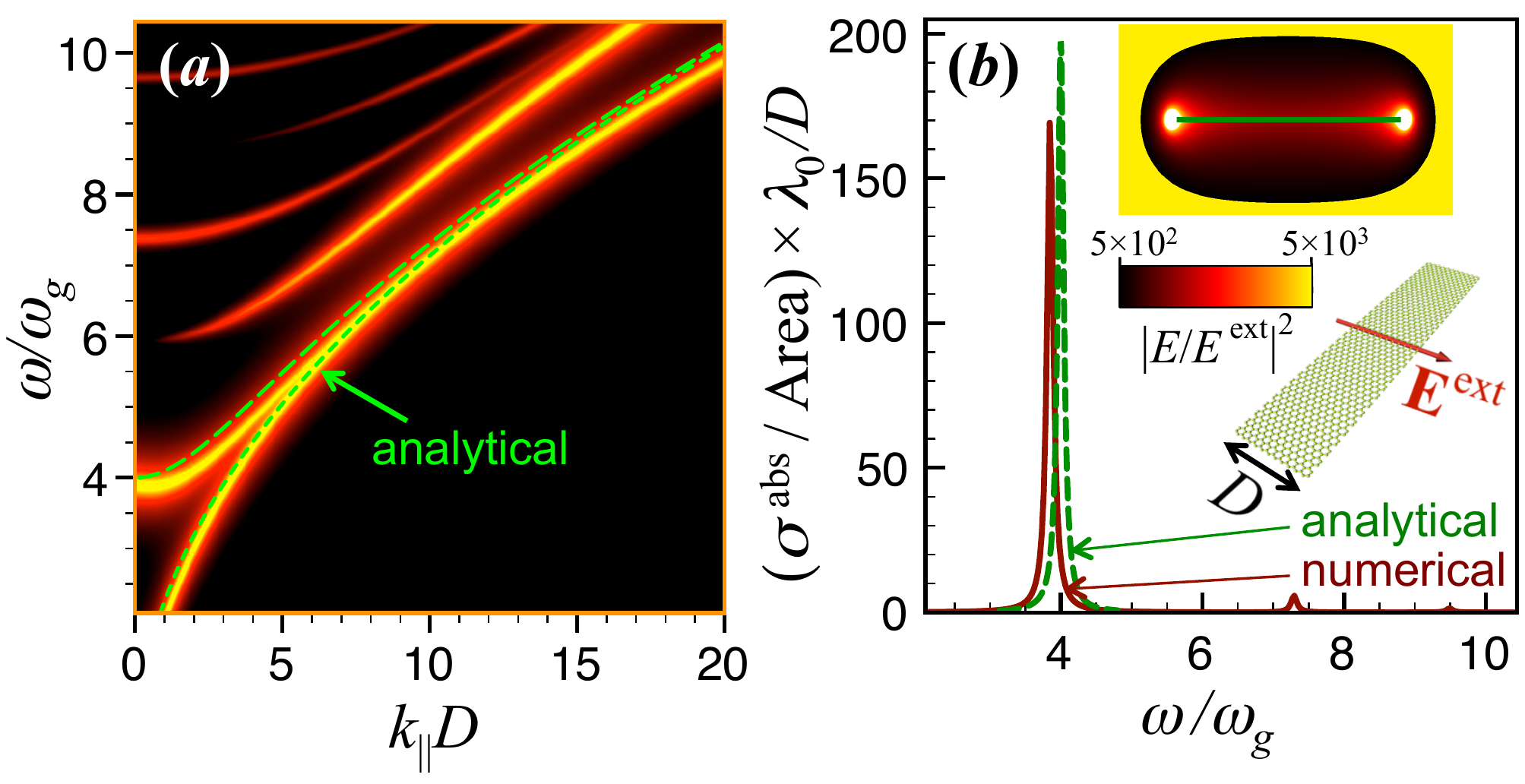}
\caption{{\bf Plasmons in individual graphene ribbons.} {\bf (a)} Absorption dispersion diagram showing the plasmons guided along a self-standing graphene ribbon of width $D$ doped to a Fermi energy $E_F$, obtained with the Drude conductivity [Eq.\ (\ref{Drude})]. The plasmon frequency is normalized to $\omega_g=(e/\hbar)\sqrt{E_F/\pi D}$. {\bf (b)} Absorption cross-section under normal incidence with light polarization as shown by the inset: numerical simulation (solid curve, see Appendix\ \ref{Ap2}) {\it vs} analytical theory [dashed curve, Eq.\ (\ref{ribbonDrudeplasmon})]. The upper inset shows the dipole-plasmon near-electric-field intensity in a plane normal to the ribbon normalized to the incident intensity for a ribbon width $D=100\,$nm. The relative intensity is only shown in the $500-5000$ range. The intrinsic width is taken as $\hbar\tau^{-1}=\omega_g/10$.} \label{Fig2}
\end{center}
\end{figure*}

\subsection{Scaling Laws for Graphene Ribbons}

Ribbons, which are central elements of many graphene plasmonics studies \cite{JGH11,paper176,NGG11,paper196,FRA12,paper181,NGG12,YLZ13,SNC13}, deserve a separate discussion. Their guided modes comprise a fundamental low-energy band of monopolar character and higher-energy modes of multipolar nature \cite{NGG11}. Using the Drude conductivity [Eq.\ (\ref{Drude})], we can obtain a universal dispersion diagram (see Fig.\ \ref{Fig2}a and Appendix\ \ref{Ap1} \cite{paper181}), in which the wave vector along the ribbon $k_\parallel$ is normalized using the width $D$.

As ribbons possess infinite area, it is convenient to write their polarizability per unit length $L$ (in the $L\rightarrow\infty$ limit) as
\begin{equation}
\frac{\alpha_\omega}{L}=D^2\sum_j\frac{A'_j}{\frac{(-2/\eta_j)}{\epsilon_1+\epsilon_2}-\frac{\ii\omega D}{\sigma(\omega)}}.
\label{alpharibbon}
\end{equation}
Similar to $A_j$ above, the new coefficients $A'_j=A_j D/L$ can be calculated once and for all by comparing this expression with a numerical solution of the polarizability (see Appendix\ \ref{Ap2}). Under normal-incidence illumination ($k_\parallel=0$), with polarization as shown by the inset of Fig.\ \ref{Fig2}b, the electrostatic polarizability of a self-standing perfect-conductor ribbon \cite{V1981}, $\alpha_0/L=D^2/16$, allows us to rewrite the sum rule of Eq.\ (\ref{rule2}) as $-\sum_j\eta_jA'_j=1/16$. Likewise, the sum rule of Eq.\ (\ref{sumAj}) now becomes $\sum_jA'_j=1$. Furthermore, the absorption spectrum is dominated by coupling to the dipolar band, so that higher-energy plasmons are hardly excited (see Fig.\ \ref{Fig2}b, solid curve). We conclude that the polarizability is dominated by the dipolar mode, with $A'_j=1$ and $\eta_j=-1/16$, according to the preceding discussion. Under this single-mode approximation, and applying the well-known expression for the absorption cross-section of a self-standing ribbon $\sigma^{\rm abs}=(8\pi^2/\lambda_0) {\rm Im}\left\{\alpha_\omega\right\}$, we find the dipole plasmon to contribute as
\begin{equation}
\frac{\sigma^{\rm abs}(\omega)}{A}=\frac{8\pi^2D}{\lambda_0}\;{\rm Im}\left\{\frac{\omega_g^2}{16\,\omega_g^2-\omega(\omega+\ii\tau^{-1})}\right\},
\label{ribbonDrudeplasmon}
\end{equation}
where $A=DL$ is the ribbon area and $\omega_g=(e/\hbar)\sqrt{E_F/\pi D}$ is a geometrical frequency. (Notice that we are actually calculating extinction, which should be nearly the same as absorption, because scattering is negligible for $D\ll\lambda_0$.) Despite its simplicity, this expression is in remarkable agreement with full numerical simulations both in the position and in the strength of the dipole plasmon (see Fig.\ \ref{Fig2}b). Incidentally, the correction produced by dielectric environment when the ribbon is placed at an $\epsilon_1|\epsilon_2$ interface shifts the plasmon energy to
\begin{equation}
E_p\approx4e\sqrt{2E_F/[\pi D(\epsilon_1+\epsilon_2)]}.
\label{ribbonwp}
\end{equation}

The field intensity is enormously enhanced near the ribbon. For $\hbar\tau^{-1}=\omega_g/10$ and $D=100\,$nm, we find a large volume with an intensity enhancement $>500$, and even $>5000$ in a region that extends up to $\sim5\,$nm away from the graphene edges (see upper inset to Fig.\ \ref{Fig2}b). It is interesting to note that the maximum intensity enhancement scales as $\tau^2 E_F/D^3$ for a fixed distance to the ribbon measured in units of the width $D$.

Guided plasmons in ribbons can be intuitively understood as the laterally confined plasmons of extended graphene, in which the in-plane wave vector has to be replaced by $\sqrt{k_\parallel^2+(m\pi/D)^2}$, where $m\pi/D$ is the transversal wave vector associated with confinement of multipole $m$. However, this procedure overlooks edge effects, which upon comparison with full numerical simulations, we find to roughly contribute to a $\sim10$\% reduction in mode frequency. With this correction, we find the analytical expression $(\omega/\omega_g)\approx0.9\sqrt{2\pi}\left((k_\parallel D)^2+m^2\pi^2\right)^{1/4}$, in reasonable agreement with the full electromagnetic simulation for the two lowest plasmon bands ($m=0,1$, see Fig.\ \ref{Fig2}a).

\subsection{Why Graphene Plasmons Enable Facile Electrical Tunability?}

The dramatic changes induced in the optical response of graphene through varying the concentration of charge carriers $n$ can be traced back to the peculiar electronic band structure of this material: because the electronic density of states vanishes at the Fermi level and the electronic bands show a linear dispersion, a relatively moderate value of $n$ produces substantial variations of $E_F\propto\sqrt{n}$ ({\it e.g.}, $E_F=0.37\,$eV for $n=10^{13}\,$cm$^{-2}$), leading to the opening of an optical gap $\sim2E_F$, where plasmons exist without undergoing Landau damping. Doping levels as high as $E_F\sim1\,$eV are currently attainable using top-gate configurations, assisted by a highly polarizable dielectric spacer \cite{CPB11}. The atomic thickness of graphene also contributes to optimize the effect of doping, in contrast to thicker materials, because the additional charge is distributed over a smaller volume.

It is useful to estimate the number of electrons participating in a plasmonic resonance via the $f$-sum rule, which has been shown to possess peculiar properties in graphene \cite{SNC08}. This allows us to quantify the above argument of electronic-density-of-states vanishing ({\it i.e.}, the fact that the number of electronic states that need to be filled by injecting electrons into the layer in order to substantially raise the Fermi energy is small compared with noble metals because the electron dispersion relation is linear and vanishes at the Fermi level of undoped graphene). In the electrostatic limit, the polarizability $\alpha_\omega$ satisfies the rigorous relation \cite{PN1966} ($f$-sum rule)
\begin{equation}
\int_0^\infty\omega\,d\omega\,{\rm Im}\left\{\alpha_\omega\right\}=\frac{\pi e^2}{2m_e}N_e,
\label{fsum}
\end{equation}
where $N_e$ is the number of electrons in the particle. In practice, we restrict the range of $\omega$ integration to the plasmonic region in order to find the effective number of electrons contributing to the plasmons. For example, in a spherical metallic particle of radius $R\ll\lambda_0$ described by the Drude permittivity $\epsilon=1-\omega_p^2/\omega(\omega+\ii\tau^{-1})$, inserting the polarizability $\alpha_\omega=R^3(\epsilon-1)/(\epsilon+2)$ into Eq.\ (\ref{fsum}), we find the relation $\omega_p^2=(3e^2/m_e)(N_e/R^3)$, which shows that $N_e$ coincides with the number of valence electrons: all valence electrons participate in the dipolar resonance of a Drude-metal sphere at frequency $\omega_p/\sqrt{3}$. For a graphene island, we use instead the polarizability given by Eq.\ (\ref{alpha}). Upon direct integration of Eq.\ (\ref{fsum}), and after using the exact relation Eq.\ (\ref{sumAj}), we find the effective density of charge carriers participating in the plasmons to be given by
\begin{equation}
n_{\rm eff}=\sqrt{n\;n_F},
\label{nelaw}
\end{equation}
where $n$ is the doping density and $n_F=(1/\pi)(mv_F/\hbar)^2\approx2.4\times10^{15}\,$cm$^{-2}$ (notice that $n_F$ has the same order of magnitude as the density of carbon atoms, $n_C=4/(3\sqrt{3}a^2)\approx3.8\times10^{15}\,$cm$^{-2}$, where $a=0.1421\,$nm is the C-C bond distance). This remarkably simple relation holds for arbitrarily shaped nanostructures in the Drude approximation [Eq.\ (\ref{Drude})]. It essentially explains that the effective density of charge carriers $n_{\rm eff}$ contributing to the plasmon is substantially higher than the doping density $n$: the effect of a relatively small number of doping electrons or holes ({\it e.g.}, one doping carrier per every 52 atoms for a Fermi energy of 1 eV) is amplified as a result of both the vanishing of the density of states at the Dirac point and the linear dispersion relation, in contrast to conventional plasmonic materials such as gold, in which the density of states takes a substantial value at the Fermi level and the outer-most s valence electrons exhibit a nearly parabolic dispersion relation.

\begin{figure*}
\begin{center}
\includegraphics[width=150mm,angle=0,clip]{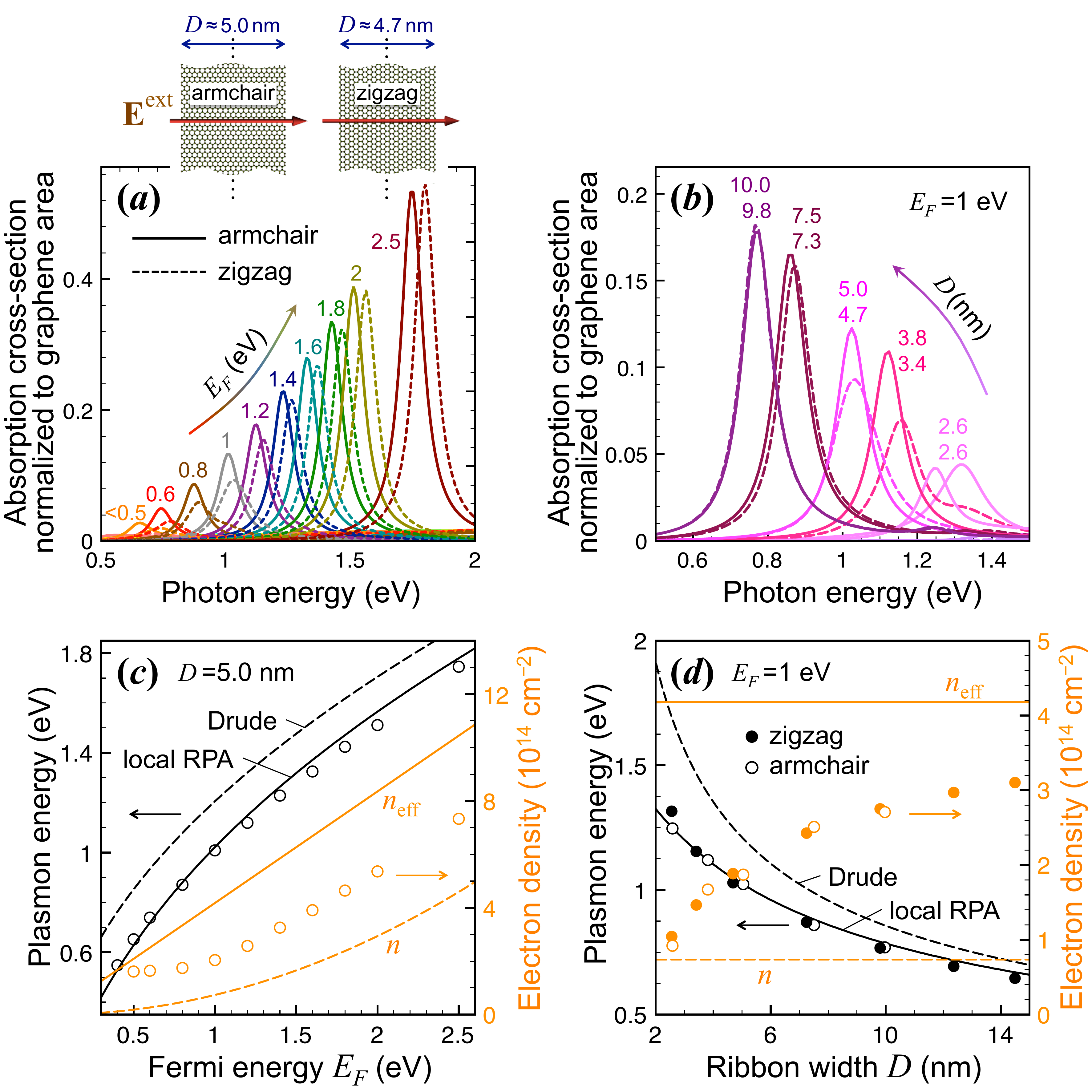}
\caption{{\bf Plasmons and optical absorption in narrow graphene ribbons.} We present absorption cross-section spectra for self-standing ribbons of both armchair and zigzag edge configurations calculated with a quantum-mechanical TB-RPA procedure, as detailed elsewhere \cite{paper183} (see Appendix\ \ref{Ap1}). {\bf (a)} Spectra of two ribbons of different edges but similar width for several doping levels. {\bf (b)} Spectra of armchair and zigzag ribbons of similar widths for $E_F=1\,$eV. {\bf (c)} Plasmon energy (left scale) and effective charge carrier density [$n_{\rm eff}$, numerically extracted from Eq.\ (\ref{nelaw}); right scale, symbols] as a function of $E_F$ for the armchair ribbon considered in (a). For reference, we also plot $n_{\rm eff}$ and $n$ for extended graphene (right scale, solid and dashed curves, respectively). {\bf (d)} Same as (c), but as a function of ribbon width $D$, as extracted from the spectra of (b). The intrinsic plasmon width is set to $\hbar\tau^{-1}=0.1\,$eV in all cases ({\it i.e.}, this is equivalent to a mobility of $66\times(E_F/{\rm eV})\;$cm$^2/$(V\,s)).} \label{Fig3}
\end{center}
\end{figure*}

We put these concepts to the test for narrow ribbons in Fig.\ \ref{Fig3}c,d (right scales), where we represent $n_{\rm eff}$ as a function of ribbon width $D$ and Fermi energy. The results of Eq.\ (\ref{nelaw}) (right scale, solid curves) are compared with values obtained from the $f$-sum rule [Eq.\ (\ref{fsum})] by integrating the plasmon peak of the spectra calculated using a quantum-mechanical approach (symbols, see Appendix\ \ref{Ap1}). We observe deviations from Eq.\ (\ref{nelaw}) that increase in magnitude with decreasing $D$. However, $n_{\rm eff}$ is larger than the carrier density $n$ (dashed curves) in all cases considered. Incidentally, the leading dipole peak in the calculated spectrum of Fig.\ \ref{Fig2}b (solid curve) accounts for $\sim85$\% of $n_{\rm eff}$ when integrated through Eq.\ (\ref{fsum}); this roughly explains a similar reduction in the observed height with respect to the analytical ribbon model (dashed curve), in which the entire plasmon weight is placed in this mode.

\section{Towards Graphene Plasmonics at Visible and Near Infrared Frequencies}

Graphene plasmons have been so far observed at mid-IR and lower frequencies \cite{FAB11,JGH11,SKK11,YLC12,YLL12,paper212,BJS13,paper196,FRA12}. It is however expected that their extension towards the vis-NIR enables unprecedentedly fast optical tunability in this spectral range, with high potential impact on telecommunications technologies. Since the plasmon frequencies scale as $\sqrt{E_F/D}$ with the Fermi energy $E_F$ and the size of the structure $D$ [see Eq.\ (\ref{wj})], an obvious way of achieving vis-NIR graphene plasmons consist in elevating $E_F$ and reducing $D$. Next, we explore some possible realizations of these prescriptions.

\subsection{Extreme Electrostatic Doping}

\begin{figure*}
\begin{center}
\includegraphics[width=170mm,angle=0,clip]{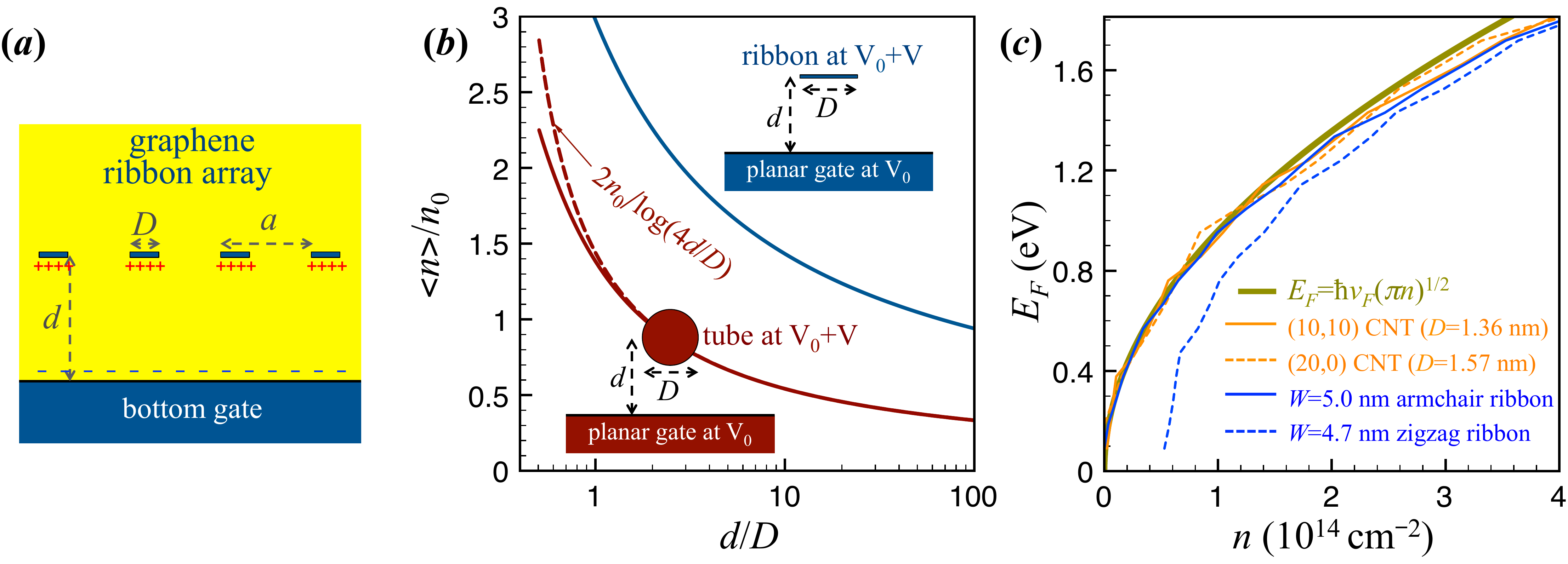}
\caption{{\bf Geometrical enhancement of electrostatic doping.} {\bf (a)} Concentration of dopping charges in a graphene layer of small filling fraction $f=D/a$. The figure represents a ribbon array that acts as a top gate and that holds the same total charge ($+$'s) as the bottom gate ($-$'s). Electrical contact to the ribbons must be provided, for example by connecting them to an extended, gated graphene island at a large distance ($\gg a,d$) from the region schematically depicted in the figure. {\bf (b)} Average charge-carrier density $\langle n\rangle$ on a planar ribbon (on the surface of a circular tube) of width (diameter) $D$ as a function of its distance $d$ to a planar gate. We normalize $\langle n\rangle$ to $n_0$, the density on each of the two plates of an infinitely extended planar capacitor separated by a distance $D$ ({\it i.e.}, $n_0=\epsilon V/(4\pi e D)$, where $\epsilon$ is the permittivity of the material mediating between all gates and $V$ is the bias potential). We compare the tube density with the exact $D/d\rightarrow0$ limit $n=2n_0/\log(4d/D)$, as obtained from the method of images (broken curve). {\bf (c)} Relation between carrier density and Fermi energy for extended graphene, as well as for ribbons and single-wall carbon nanotubes (CNTs). The latter are extracted from a TB description of the electronic bands (see Appendix\ \ref{Ap1}).} \label{Fig4}
\end{center}
\end{figure*}

In typical electrostatic gating configurations, the graphene accumulates charge carriers by acting as one of the two plates of a capacitor, separated from the other plate by a dielectric. A simple method to increase the doping density consists in patterning the graphene with a small in-plane carbon filling fraction $f$, as shown in Fig.\ \ref{Fig4}a. For a periodic pattern of small period compared with the distance to the other gate, the average charge density in the graphene plane should be roughly independent of $f$, but as this charge is only supported by the conducting carbon layer, the doping density in the graphene becomes $n/f$, where $n$ is the density obtained with an unpatterned gate. Realistically, we can consider an array of lithographically patterned ribbons of width $D=20\,$nm, spaced with a period of 200\,nm ({\it i.e.}, $f=0.1$), and placed at a distance $d$ of a few hundred nanometers from a planar gate. Considering that Fermi energies as high as $E_F\approx1\,$eV have been achieved using top-gate electrical doping in extended graphene \cite{CPB11}, this strategy would boost the Fermi energy to $E_F/\sqrt{f}\sim3.2\,$eV. Using the above scaling law for ribbons  [Eq.\ (\ref{ribbonwp})], and assuming the graphene to be supported on an $\epsilon=2$ substrate, we find a plasmon at a NIR photon wavelength $\sim1.4\,\mu$m. Incidentally, at this high level of doping, the nonlinearity of the electronic band structure becomes relevant.

Additionally, this approach could be reinforced by operating at a high chemical-doping base point, whereby the graphene Fermi energy could be doped above 1\,eV in the absence of any bias \cite{KWB12}, while gating would be used to modulate $E_F$ around this point.

In a similar fashion, the average doping density of an individual graphene ribbon in the $d\gg D$ limit is $\propto n_0/\log(d/D)$ \cite{paper194}, where $n_0=\epsilon V/(4\pi eD)$ is the density in a planar capacitor with a separation $D$ between the gates. In this limit, the level of doping is controlled by the ribbon width $D$, so that high densities can be achieved without suffering electrical breakdown because the ribbon and the planar gate are separated by a large distance $d$. We illustrate this possibility in Fig.\ \ref{Fig4}b for both ribbons and tubes. At a distance $d\sim100\,D$, the doping density still reaches values comparable to $\sim n_0$. Moreover, as the electric field is high in the proximity of the ribbon (tube), it must be weaker near the planar gate compared with the uniform field inside a planar capacitor for the same bias potential $V$ and separation $d$ between the gates. Therefore, one would expect that $V$ could be raised without causing breakdown to at least a similar level as for a conventional planar capacitor of separation $d$. This results in a $d/D-$fold increase in doping density of the ribbon (tube), or equivalentely, a $\sqrt{d/D}-$fold increase in $E_F$. Fermi energies as high as a few electronvolts seem to be within reach following this prescription. Incidentally, the scaling $E_F\propto\sqrt{n}$ works reasonably well for $E_F<2\,$eV with D down to the nanometer range, even when a proper account of electron bands is taken into consideration in single-layer ribbons or single-wall tubes (see Fig.\ \ref{Fig4}c).

An intriguing situation might be encountered when the doping charge per carbon atom is comparable to unity. In particular, considering the results of Fig.\ \ref{Fig4}b, for a $D=10\,$nm ribbon embedded in an $\epsilon=10$ dielectric at a distance $d=1\,\mu$m from a planar gate with a 500\,V bias relative to the ribbon, the doping charge reaches a value of $0.24$ carriers per carbon atom. A nonlinear electrostatic regime is then achieved, with is made even more dramatic due to the $1/\sqrt{x}$ divergence of the doping charge with decreasing distance $x$ to the graphene edges \cite{paper194}.

\subsection{Plasmons in Narrow Ribbons and the Effect of Edge Damping}

In a recent study, plasmons in graphene nanorings were observed down to a wavelength of $\sim3.7\,\mu$m \cite{paper212}. The rings were doped to $E_F\approx0.8\,$eV and placed in an $\epsilon\sim2$ environment. More precisely, this was the anti-bonding plasmon mode, which shows up at an energy similar to the dipole plasmon of a ribbon with similar width [Eq.\ (\ref{ribbonwp})]. This required to cut the ring with a challenging $\sim20\,$nm width using electron-beam lithography \cite{paper212}. Narrower graphene ribbons, which can be grown on vicinal surfaces \cite{HTT13}, selected using colloid chemistry methods \cite{LWZ08}, and also synthesized via self-assembly of organic molecules \cite{CRJ10,LTA13}, constitute a natural way of pushing plasmons further towards the vis-NIR using a scalable bottom-up approach.

For such small sizes, the orientation of ribbon edges becomes critical because zigzag edges can broaden the plasmons enormously, as shown through quantum-mechanical simulations \cite{paper183}. Edge damping is thought to be caused by the presence of zero-energy electronic edge states, and it is particularly active when the plasmon energy $E_p$ is above $E_F$, so that decay through excitation of those states becomes physically possible. If zigzag and armchair edges are mixed, such as in graphene disks, edge damping takes on even for $E_p<E_F$ \cite{paper183}, as high-momentum transfers are then favored, thus reducing the effective size of the optical gap (see Fig.\ \ref{Fig1}f). However, ribbons with uniform edges constitute clean systems on which edge damping can be prevented by having $E_p<E_F$.

Figure\ \ref{Fig3}a shows that modulation of the graphene plasmon energy $E_p$ is possible in both armchair and zigzag $\sim5\,$nm ribbons. For $E_F>1\,$eV, we have $E_p<E_F$, and as anticipated, edge effects play a minor role, so that both types of ribbons produce similarly narrow and intense absorption features. By contrast, for $E_F<1\,$eV, edge damping switches on in zigzag ribbons. A similar conclusion can be extracted upon inspection of ribbons of different widths (see Fig.\ \ref{Fig3}b): when $E_p$ is pushed above $E_F=1\,$eV by narrowing the ribbon width, zigzag ribbons produce less intense plasmons.

Overall, we conclude that plasmons in narrow graphene ribbons can be modulated at vis-NIR frequencies and produce absorption cross-sections comparable to the graphene area. Incidentally, we have used a very conservative estimate of the intrinsic width in Fig.\ \ref{Fig3}, corresponding to a mobility of only $\mu=66\times(E_F/{\rm eV})\;$cm$^2/$(V\,s)). As the height of the absorption peaks is proportional $1/\mu$, much higher cross-sections are expected in practice, possibly reaching several times the graphene area.

\subsection{Towards Molecular Plasmonics}

Graphene quantum dots with diameters $D<10\,$nm and below have been synthesized by resorting on chemical methods rather than lithography \cite{KHK12}. Now, assuming a doping level $E_F=1\,$eV and an $\epsilon=2$ host medium, these structures should sustain plasmons down to $\sim700\,$nm light wavelength, according to the above scaling law. For such small diameters, finite size and edge effects are predicted to dramatically damp the plasmons \cite{paper183}, unless the dot edges are preferentially armchair, rather than zigzag \cite{paper214}. However, control over edge orientations is rather limited, and so is size selection through filtering and dialysis. A similar lack of control over edges is encountered in nanometer-sized graphene quantum dots obtained from fullerenes \cite{LYG11}.

Alternatively, chemical synthesis allows producing large polycyclic aromatic hydrocarbons (PAHs) \cite{WPM07}, which can be regarded as truly nanometer-sized graphene islands with better control over size, edge orientation, and morphology. These molecules have been recently postulated as a viable alternative to engineer tunable devices in the vis-NIR spectral region \cite{paper215}: their collective electron excitations are of similar nature as graphene plasmons and can equally be tuned through electrical gating. The plasmonic character of these excitations is revealed by the important role played by electron-electron interactions \cite{paper215}. Nonetheless, gating these molecules constitutes an experimental challenge that must be overcome in order to achieve electro-optical tunability.

\begin{figure*}
\begin{center}
\includegraphics[width=140mm,angle=0,clip]{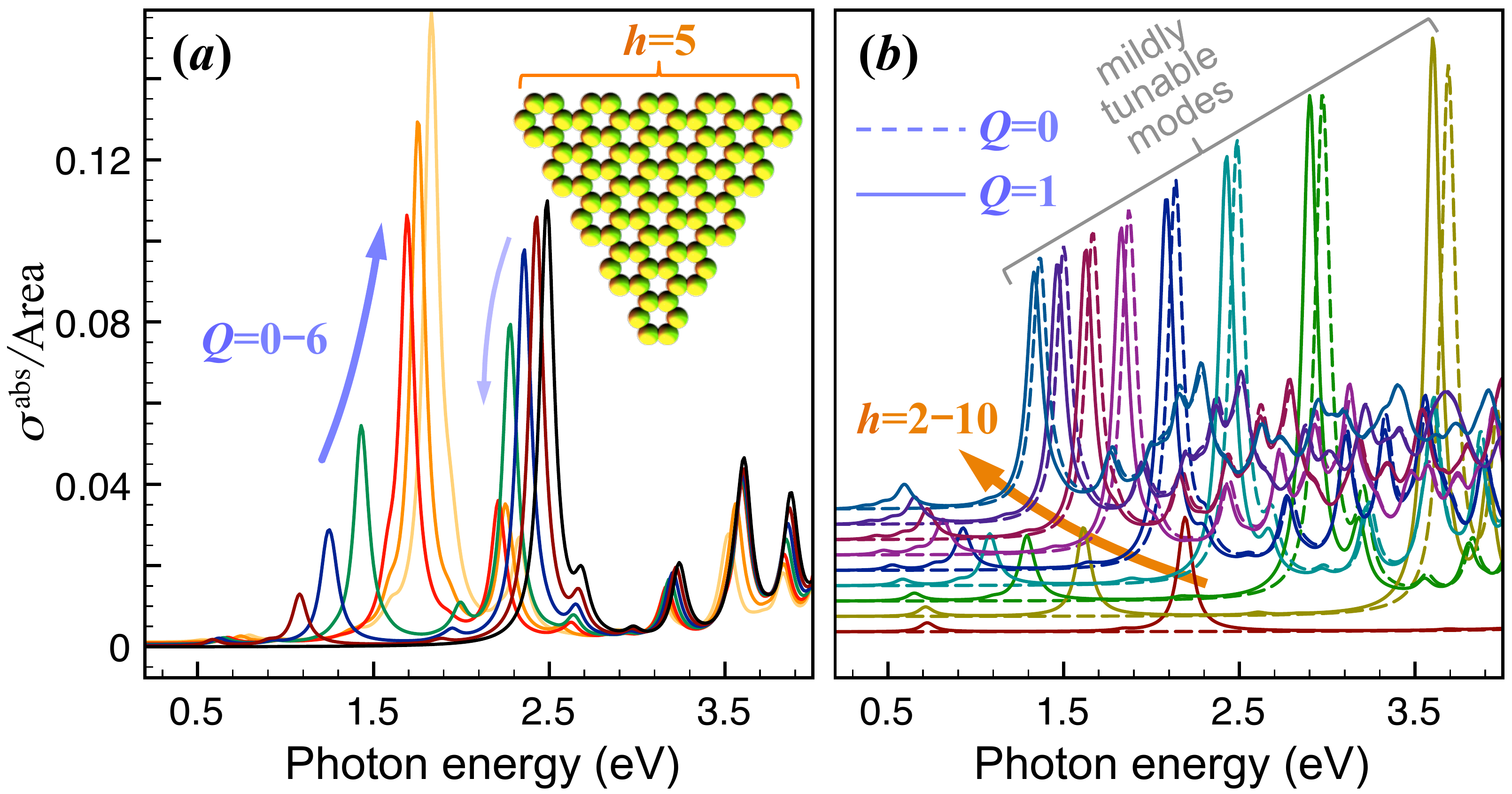}
\caption{{\bf Molecular plasmons in nanographene.} {\bf (a)} Absorption by an armchair graphene triangle spanning $h=5$ hexagons per side and with different numbers of charge carriers $Q=0-6$. {\bf (b)} Same as (a) for $h=2$ (tryphenylene) to $h=10$ in neutral ($Q=0$, dashed curves) or singly charged ($Q=1$, solid curves) states. Spectra corresponding to different $h$'s have been vertical offset for clarity. The absorption cross-section is normalized to the graphene area ($3\sqrt{3}Na^2/4$, where $N=3h(h+1)$ is the number of carbon atoms and $a$ is the C-C bond distance). All calculations are performed with the TB+RPA approach (see Appendix\ \ref{Ap1} \cite{paper183}) assuming an intrinsic damping $\hbar\tau^{-1}=0.1\,$eV.} \label{Fig5}
\end{center}
\end{figure*}

Following the ideas introduced in Ref.\ \cite{paper215}, we analyze in Fig.\ \ref{Fig5} the low-energy optical resonances of small armchair graphene nanoislands in neutral and ionized states. The spectra are calculated using the TB+RPA approach (see Appendix\ \ref{Ap1}), which produces results in qualitative agreement with first-principles simulations \cite{paper215}. The 90-carbon-atom nanotriangle considered in Fig.\ \ref{Fig5}a exhibits a $\sim2.5\,$eV absorption gap in its neutral state ($Q=0$). However, when ionized, an absorption feature emerges in this gap, which evolves towards higher energies as the number $Q$ of additional electrons or holes increases, just like we have predicted from classical theory for larger structures. This evolution is accompanied by a much weaker modulation of the second absorption peak, which is however moving towards lower energies with increasing $Q$. We also find that the magnitude of the optical gap decreases with increasing triangle size (see Fig.\ \ref{Fig5}b, showing spectra for structures ranging from triphenylene, consisting of 18 carbon atoms, to a 330 carbon-atom triangle), while the lowest-energy plasmon experiences a redshift also similar to plasmons in larger graphene islands. A similar evolution with size and $Q$ is observed when considering larger structures, which are quantitatively describable through classical theory above $\sim10\,$nm in size \cite{paper183}, thus making a smooth transition between the regimes of molecular excitations and graphene plasmons. From a practical viewpoint, the switching on and off of this low-energy feature from the neutral to the singly ionized state constitutes a potentially viable approach to achieve optical modulation in the vis-NIR \cite{paper215}.

\begin{figure*}
\begin{center}
\includegraphics[width=160mm,angle=0,clip]{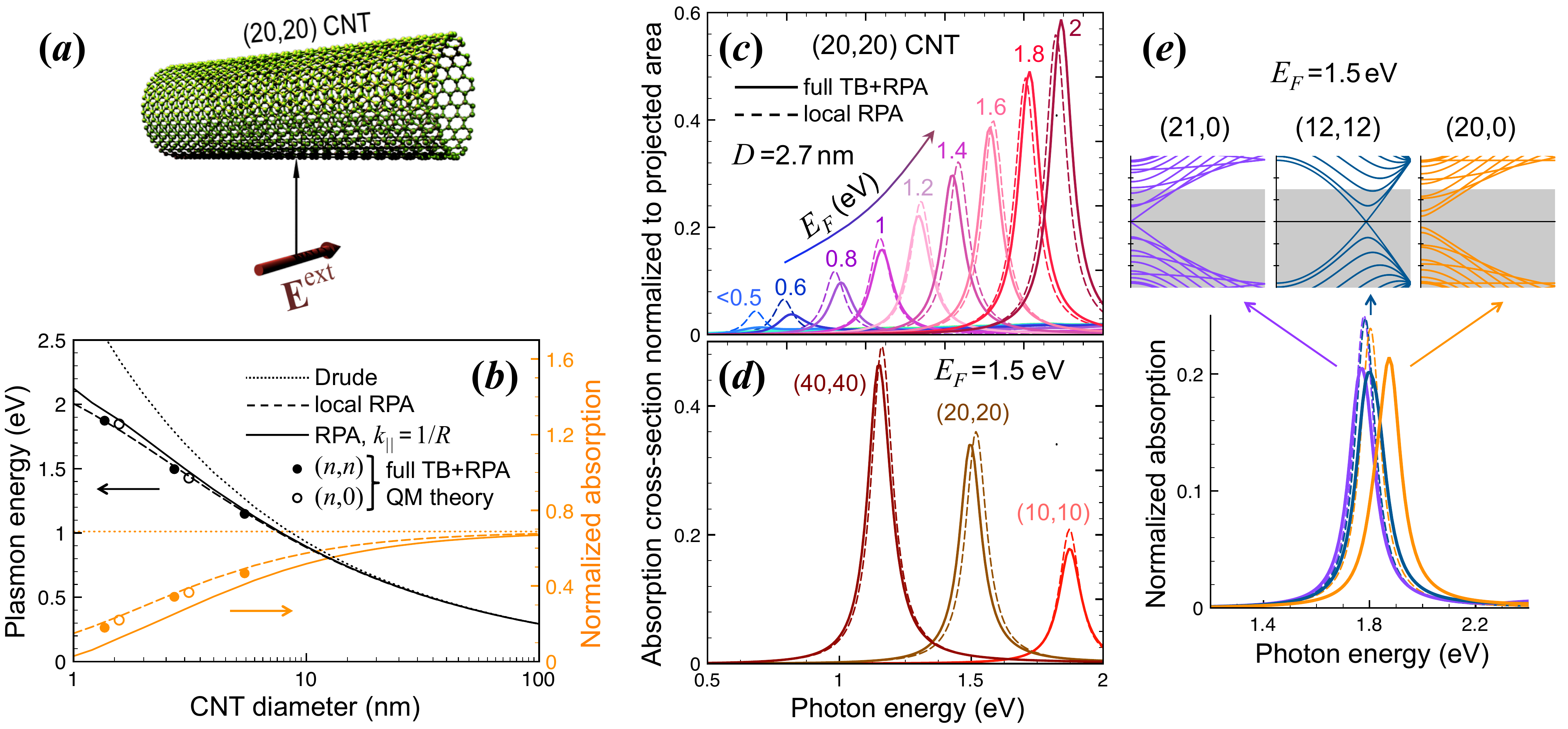}
\caption{{\bf Tunable vis-NIR plasmons in doped carbon nanotubes.} {\bf (a)} Single-wall carbon nanotube (CNT) subject to external illumination with polarization perpendicular to the tube. {\bf (b)} Plasmon energy (left scale) and peak absorption (right scale) as a function of CNT diameter. {\bf (c)} Absorption spectra for a (20,20) tube (diameter $D=2.7\,$nm) doped to different Fermi levels $E_F$. {\bf (d)} Absorption spectra for armchair CNTs of different size and $E_F=1.5\,$eV. {\bf (e)} Absorption spectra of three different CNTs of similar diameter but different electronic structure (see electronic bands over the first Brillouin zone in the upper insets, where the occupied levels are shown by a shaded area up to an energy $E_F=1.5\,$eV). The absorption cross-section is normalized to the projected area of the tubes, while the intrinsic optical width is taken as $\hbar\tau^{-1}=0.1\,$eV in all cases. Full TB+RPA quantum mechanical simulations (solid curves, see Appendix\ \ref{Ap1}) are compared with analytical local-RPA theory [broken curves, Eq.\ (\ref{sigmaCNT})] in (c-d).} \label{Fig6}
\end{center}
\end{figure*}

\subsection{Plasmons in Doped Carbon Nanotubes}

Carbon nanotubes (CNTs) are free from the damaging effects produced by edges, thus offering an attractive alternative to obtain high-energy tunable plasmons. For sufficiently large diameter $D$, the optical properties of a single-wall CNT should be describable as those of a circular cylinder characterized by the same surface conductivity $\sigma$ as planar graphene, because the effects of curvature along the perimeter are expected to play a minor role in that limit. Focusing on external light polarized across the CNT (Fig.\ \ref{Fig6}a), direct solution of Poisson's equation yields (see Appendix\ \ref{Ap3})
\begin{align}
\frac{\sigma^{\rm abs}(\omega)}{\rm area}
&=\frac{\pi^2D}{\lambda_0}\;{\rm Im}\left\{\frac{-1}{1+4\pi\ii\sigma/\omega D}\right\} \nonumber\\
&=\frac{4\pi^3D}{\lambda_0}\;{\rm Im}\left\{\frac{\omega_g^2}{4\pi\omega_g^2-\omega(\omega+\ii\tau^{-1})}\right\}
\label{sigmaCNT}
\end{align}
for the absorption cross-section normalized to the projected tube area. The last expression in \ref{sigmaCNT} is obtained in the Drude model for $\sigma$, to be compared with the above analogous result for ribbons (\ref{ribbonDrudeplasmon}). We thus conclude that CNTs present similar tunability properties as ribbons of comparable width $D$, with the plasmon frequency shifted from $4\omega_g$ in ribbons (\ref{ribbonDrudeplasmon}) to $2\sqrt{\pi}\omega_g$ in CNTs (\ref{sigmaCNT}), where $\omega_g=(e/\hbar)\sqrt{E_F/\pi D}$.

The prediction of \ref{sigmaCNT} is remarkably close to full quantum-mechanical TB+RPA  theory (see Appendix\ \ref{Ap1} \cite{paper183}), as Fig.\ \ref{Fig6} demonstrates. In particular, by inserting Eq.\ (\ref{local}) ({\it i.e.}, the local-RPA model for $\sigma$) into Eq.\ (\ref{sigmaCNT}), we find extinction spectra (Fig.\ \ref{Fig6}c, dashed curves) in excellent agreement will quantum-mechanical TB+RPA calculations (Fig.\ \ref{Fig6}d, solid curves), except at low doping levels, for which local-RPA produces larger cross sections because it does not take into account the full nonlocal dependence of electron-hole-pair excitations. Similar to ribbons, we find these additional damping effects to be active for plasmon energies $E_p>E_F$.

In Fig.\ \ref{Fig6}c, we analyze the evolution of absorption spectra as a function of doping for a (20,20) CNT, while Fig.\ \ref{Fig6}d illustrates the size dependence for fixed doping. Despite the large intrinsic damping that is assumed ($\hbar\tau^{-1}=0.1\,$eV), we obtain absorption cross-sections comparable to the projected area of the tubes. As predicted above, the plasmon frequency  exhibits a $\propto\sqrt{E_F/D}$ behavior, while the peak cross-section increases both with $E_F$ and with $D$ (see Fig.\ \ref{Fig6}b).

A drawback of CNTs for electronics applications is the difficulty in synthesizing large quantities of them with the same chirality $(n,m)$. The tubes are metallic or semi-conducting depending on the value of $n-m$. By contrast, the plasmonic properties do not seem to depend so much on chirality: plasmons are roughly controlled by the average electron density when their energies are well above the gaps. Indeed, we find that tubes of approximately the same diameter but different chirality feature plasmons of similar energy and strength (see lower part of Fig.\ \ref{Fig6}e), despite their very different band structures (upper insets). In particular, we compare in Fig.\ \ref{Fig6}e the absorption spectra of a small-gap semi-conducting (21,0) tube, a metallic (12,12) tube, and a moderately semiconducting (20,0) tube (diameters$\,\sim1.6\,$nm). Local-RPA dielectric theory (dashed curves) produces similar results for all three of them, although more realistic TB+RPA simulations reveal a blue shift in the semiconducting tube. Despite the demonstrated ability to perform spectroscopy on individual CNTs \cite{SEW06}, including the absolute determination of absorption cross-sections \cite{BPT13,LHC14}, CNT ensembles should be simpler to integrate in actual devices \cite{PKS13}. In this respect, the present results, and in particular the small dependence of the plasmons on chirality, provide a solid basis to postulate size-selected CNTs as a viable platform for light modulation in the vis-NIR.

\section{Further Directions}

\subsection{Perfect Tunable Optical Absorption}

Prospects for several promising optical applications of graphene, including light harvesting, spectral photometry \cite{FLW12,VVC12}, and optical modulation \cite{LYU11}, rely on achieving a high level of absorption by the single-atom carbon layer. Unfortunately, undoped graphene is a poor absorber, characterized by a nearly constant absorbance roughly equal to $\pi\alpha\approx2.3\%$ \cite{MSW08}. Nanostructured undoped graphene absorbs even more poorly, as shown by Eq.\ (\ref{undoped}) (see also Fig.\ \ref{Fig8}c). Graphene plasmons provide a way of enhancing absorption, with the additional advantage of being electrically tunable, so that the spectral region of high absorbance can be scanned over the range of interest.

Complete optical absorption has been predicted for periodic arrays of graphene disks under the condition that the absorption cross-section of each disk is comparable to the unit cell area \cite{paper182}. Progress towards the implementaion of this concept has been made through experiments showing electro-optical modulation of plasmonic absorption in disk and ring arrays in excellent agreement with theory \cite{paper212}, while $>30$\% measured absorption has been recently reported \cite{paper230}.

\begin{figure*}
\begin{center}
\includegraphics[width=140mm,angle=0,clip]{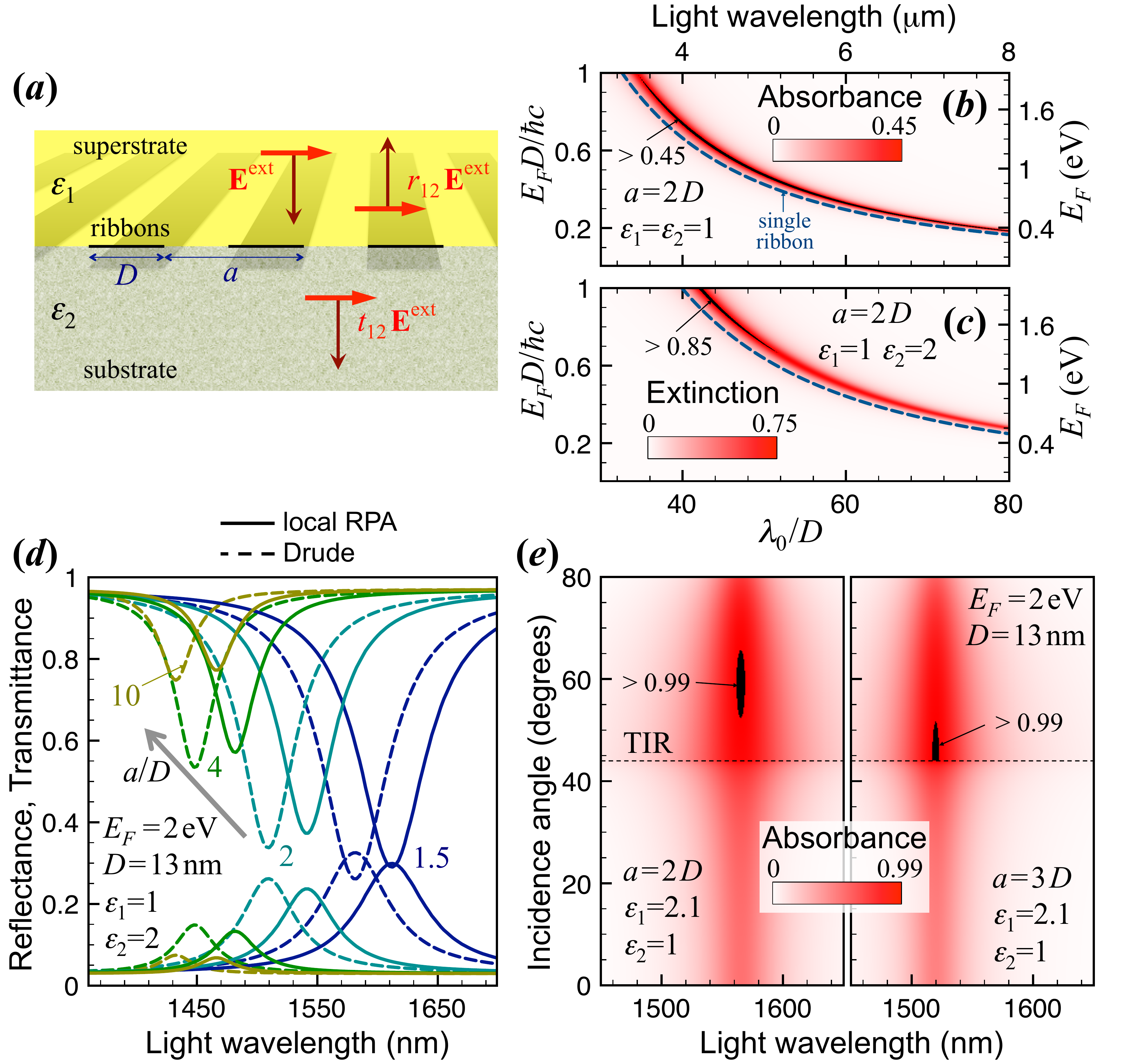}
\caption{{\bf Enhanced optical absorption in periodic arrays of ribbons.} {\bf (a)} Scheme of a ribbon array sandwiched between media of permittivities $\epsilon_1$ and $\epsilon_2$. The array is subject to normal-incidence illumination with polarization across the ribbons. {\bf (b)} Absorbance by a self-standing array ($\epsilon_1=\epsilon_2=1$) as a function of normalized light wavelength and Fermi energy. {\bf (c)}  Extinction ($=1-$transmittance) of a supported array ($\epsilon_1=1$, $\epsilon_2=2$). The Drude conductivity is assumed in both (c) and (d), where the dashed curves show the plasmon dispersion of an individual ribbon, while the right and upper scales correspond to a specific ribbon width $D=100\,$nm. {\bf (d)}  Reflectance (lower curves) and transmittance for arrays of $D=13\,$nm supported ribbons with different periods $a$ (see labels) and two different models for the conductivity. {\bf (e)}  Absorption for the same ribbons as in (d), supported on glass and illuminated from the glass side. Regions of total absorption are signaled in black above the threshold for total internal reflection (TIR). Light is incident in the plane parallel to the ribbons and perpendicular to the glass surface, with the electric field oriented across the ribbons. All calculations are based on a dipole model for the ribbons [see Eq.\ (\ref{Aarray})], assuming a quality factor of 50.} \label{Fig7}
\end{center}
\end{figure*}

The absorption of arrayed graphene ribbons has been studied in detail following modal expansions \cite{BPV12,FP12} and finite-difference \cite{NGG12} computation methods. Here, we discuss absorption in these structures through analytical methods, which combine the results discussed above for individual ribbons with the methods developed to analytically investigate similar phenomena in 2D arrays of finite graphene islands \cite{paper182,paper212}. In particular, we discuss perfect absorption in arrays of graphene ribbons periodically arranged with period $a$ along the interface between two media of real refractive indices $n_1=\sqrt{\epsilon_1}$ and $n_2=\sqrt{\epsilon_2}$. This analysis can be straightforwardly applied to CNTs as well. For simplicity, we focus first on normally incident light coming from medium 1 and polarized across the ribbons, as shown in Fig.\ \ref{Fig7}a. The ribbons are described through their polarizability per unit length $\alpha_\omega/L$ [see Eq.\ (\ref{alpharibbon})]. Following a similar approach as in previous studies \cite{paper182,paper212}, we find the following exact result (within the dipole model) for the transmission and reflection coefficients (see Fig.\ \ref{Fig7}a):
\begin{align}
t_{12}&=\frac{2n_1}{n_1+n_2}\left(1+\frac{\ii S}{(\alpha_\omega/L)^{-1}-G}\right),\nonumber \\  
r_{12}&=t_{12}-1,
\label{r12}
\end{align}
where $S=(4\pi^2/a\lambda_0)\,[2/(n_1+n_2)]$ and $G=(g/a^2)\,[2/(\epsilon_1+\epsilon_2)]+\ii S$, whereas $g$ represents a lattice sum over dipole-dipole interactions, which in the long-wavelength limit ($\lambda_0\gg a$) reduces to $g=2\pi^2/3$. From here, we obtain the absorbance as
\begin{equation}
{\rm Absorbance}\,=1-|r_{12}|^2-\frac{n_2}{n_1}|t_{12}|^2.
\label{Aarray}
\end{equation}
In what follows, we use the single-mode approximation for the graphene polarizability  $\alpha_\omega/L=D^2/(16-\ii\omega D/\sigma)$ [see discussion of Eq.\ (\ref{ribbonDrudeplasmon}) above], with $\sigma$ calculated from different models (see captions and labels in Figs.\ \ref{Fig7} and \ref{Fig8}).

Obviously, the absorbance is enhanced near the frequency of the individual ribbon plasmon (Fig.\ \ref{Fig7}b,c, dashed curves), although interaction across the lattice produces a significant redshift. Inserting Eq.\ (\ref{r12}) into Eq.\ (\ref{Aarray}), the absorbance becomes a function of the complex variable $t_{12}$, which takes a maximum value of $n_1/(n_1+n_2)$ under normal incidence from medium 1. This value can be reached under so-called critical-coupling conditions \cite{paper182}, which are actually produced with graphene ribbons at a point within the highlighted black region of high absorbance in Fig.\ \ref{Fig7}b. Incidentally, inclusion of interband transitions produces a redshift ({\it cf.} dashed and solid curves in Fig.\ \ref{Fig7}d), in agreement with recently reported hydrodynamic simulations \cite{WK13}. In a more realistic configuration, with graphene supported on a substrate, large extinction ($=1-\,$transmittance$\,=1-(n_2/n_1)|t_{12}|^2$) is also predicted (see Fig.\ \ref{Fig7}c), down to NIR wavelengths for narrow ribbons (Fig.\ \ref{Fig7}d).

The conditions for which total absorption can be obtained through plasmon excitation in arrays of graphene islands have been identified in a previous study \cite{paper182}, illustrated by examples based upon the so-called Salisbury screen configuration. Here, we present further results of total absorption, using ribbons instead of finite islands. In particular, we predict this effect to take place under total internal reflection conditions (see Fig.\ \ref{Fig7}e, calculated from an extension of the above expressions to oblique incidence, as described in Appendix\ \ref{Ap4}), for which the transmission channel is already suppressed. We observe total absorption for a wide range of spacing parameters $a$, thus indicating that the effect is robust.

\subsection{Ultrafast Graphene Optics: Transient Plasmons}

Transient plasmons produced by optical heating in graphene islands may provide a viable solution to extend the plasmonic response of this material to the vis-NIR. The idea is as follows: (i) an off-resonance femtosecond pulse can be used to optically pump the carbon layer, thus creating a heated valence electron gas, which typically takes a few tens of femtoseconds to reach thermal equilibrium at a temperature as high as several thousand degrees, followed by slow heat diffusion through the thermal conductivity of the surrounding materials \cite{BTM13,TSJ13,JUN13,GSW13}; (ii) during the sub-picosecond time window over which the electron gas is at an elevated temperature, a second spectrally tuned probing pulse can excite plasmons of similar nature as the thermoplasmons predicted for extended undoped graphene \cite{V06_2}. We should note that transient optical effects have been extensively studied in the past, including the ultrafast dynamics of plasmons in nanoparticles \cite{PBL97}, the metallic behavior in optically pumped semiconductors \cite{SL00}, the transient absorption of molecules \cite{BGT09} and graphene \cite{GSW13}, and the nonlinear refractive index of graphene oxide \cite{YLZ13_1}. We concentrate here on transient plasmons in graphene, which offer a unique opportunity because of the relatively low electron heat capacity of this material ({\it i.e.}, as a result of the Dirac-cone electronic structure, a realistic pulse fluence can produce extremely high electron temperatures, as we discuss below).

\begin{figure*}
\begin{center}
\includegraphics[width=160mm,angle=0,clip]{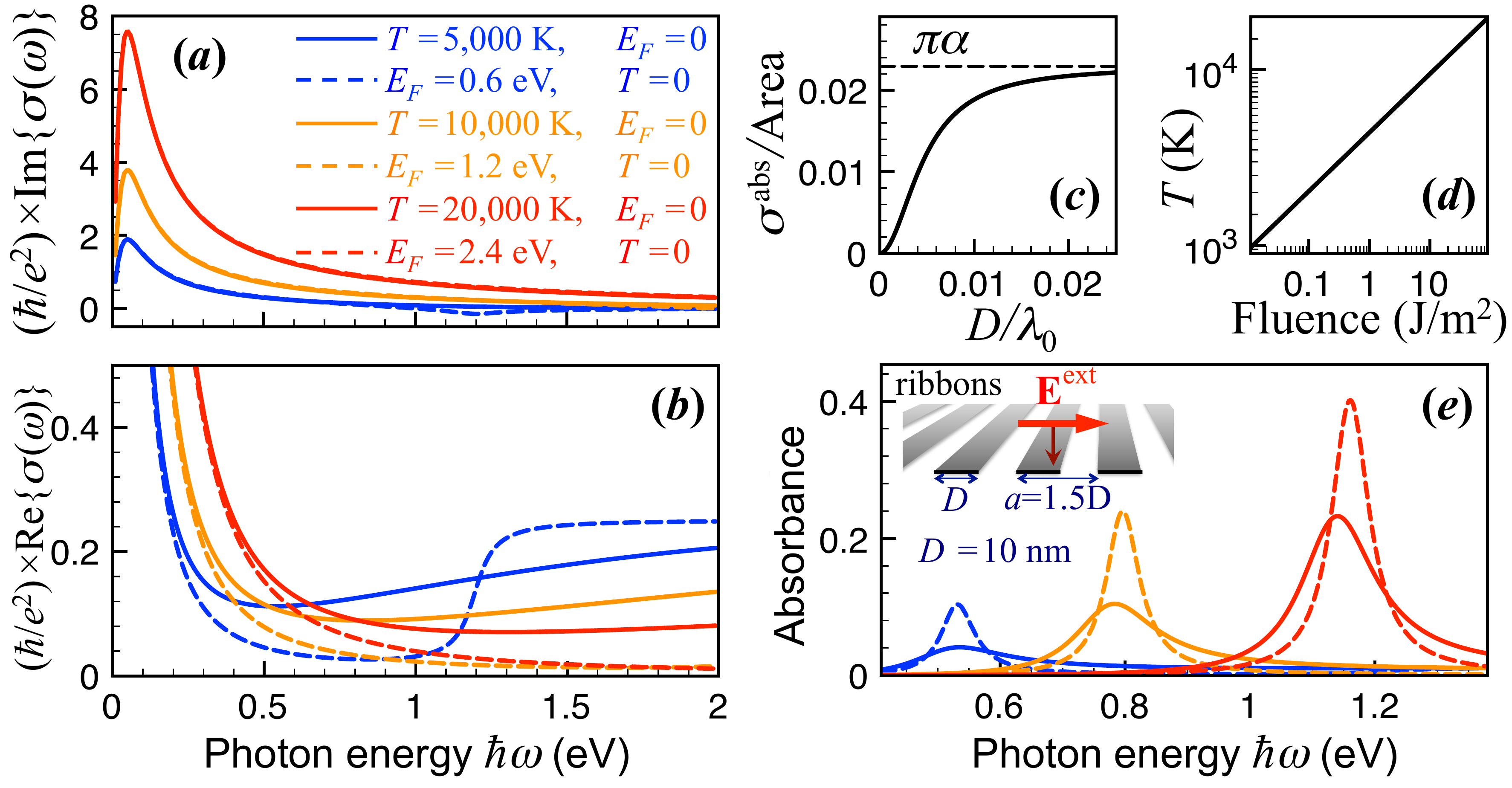}
\caption{{\bf Thermally activated transient plasmons.} {\bf (a,b)} Real and imaginary parts of the conductivity of graphene for different temperatures and doping levels (see labels in (a)). {\bf (c)} Absorption cross-section normalized to graphene area for undoped ribbons as a function of the width to light-wavelength ratio $D/\lambda_0$ [see Eq.\ (\ref{undoped})]. {\bf (d)} Electron temperature reached by undoped extended graphene as a function of laser pulse fluence. {\bf (e)} Absorbance spectra obtained from Eq.\ (\ref{Aarray}) for a self-standing array of $D=10\,$nm graphene ribbons under the same doping and temperature conditions as in (a). We use the local-RPA conductivity [{\it i.e.}, Eq.\ (\ref{local})] with an intrinsic damping $\hbar\tau^{-1}=0.05\,$eV in all cases ({\it i.e.}, this is equivalent to a mobility of $132\times(E_F/{\rm eV})\;$cm$^2/$(V\,s)).} \label{Fig8}
\end{center}
\end{figure*}

A heating laser pulse of fluence $F$ transfers an energy $Q=\sigma^{\rm abs}F$ to the graphene electrons. Now, the relation between $Q$ and the electron-gas temperature $T$ at thermal equilibrium can be worked out from the electron dispersion relation, which we assume to be the same as in extended graphene for an order-of-magnitude estimate of the $Q(T)$ function. Considering that the electron (and hole) energies involved are close enough to the Dirac point as to assume a linear dispersion relation ({\it i.e.}, $k_BT<2\,$eV), and further assuming an equilibrium Fermi-Dirac distribution of electron (hole) energies, we find $Q=s A(k_BT)^3/(\hbar v_F)^2$, where $s=(2/\pi)\int_0^\infty\theta^2d\theta/(1+\ee^\theta)\approx1.15$ and $A$ is the graphene area. This result coincides with the in-plane thermal electron energy of graphite \cite{KN1951,NI03}, which should be a reasonable approximation for high $T$. From this analysis, assuming $\sigma^{\rm abs}=\pi\alpha A$ (see Fig.\ \ref{Fig8}c), we find a temperature $T=10^4\,$K with $F\approx12\,$J$/$m$^2$ (see Fig.\ \ref{Fig8}d), which is a level of fluence commonly used in  ultrafast experiments. It should be noted that, although the electron gas reaches a high temperature, the carbon lattice has a much higher thermal capacity, and consequently, the entire system ends up at a substantially lower temperature at thermal equilibrium. Using measured data for the heat capacity of graphite \cite{PVR12}, the relaxation of the $10^4\,$K electron gas is estimated to produce just a 60\,K increase in the lattice temperature starting from ambient conditions. Therefore, heating damage should be negligible. We have obviously neglected diffusive and radiative cooling, which should play a relatively small role over a sub-picosecond time window following electron thermalization. Also, we have ignored optical phonons, which couple strongly to hot electrons and holes \cite{BTM13} and need to be included as a factor that reduces the energy deposited on the thermalized valence band; however, this factor can be easily compensated by increasing the pumping intensity. Incidentally, electrons and holes provide similar contributions to the optical response of the heated valence band, due to the symmetry of the Dirac band structure in graphene.

The intraband term of the local conductivity [Eq.\ (\ref{local})] can be written in closed-form when the electron gas is at thermal equilibrium with a Fermi-Dirac energy distribution $f_E=1/[1+\ee^{(E-E_F^0)/k_BT}]$ at temperature $T$ around a relaxed Fermi energy $E_F^0$. It just reduces to the Drude conductivity of Eq.\ (\ref{Drude}), but with $E_F$ substituted by a temperature-corrected Fermi energy \cite{FP07_2}
\begin{equation}
E_F=E_F^0+2k_BT\,\log\left(1+\ee^{-E_F^0/k_BT}\right).
\label{EFT}
\end{equation}
The increase in Fermi energy can be substantial for large $T$ in undoped graphene ($E_F^0=0$). The Drude conductivity corresponding to values of $E_F$ obtained from Eq.\ (\ref{EFT}) for temperatures $T=5-20\times10^3\,$K (or equivalently, $k_BT=0.43-1.72\,$eV) is represented in Fig.\ \ref{Fig8}a,b (dashed curves) and compared with the full local-RPA conductivity [solid curves, calculated from Eq.\ (\ref{local}) for finite $T$ and $E_F^0=0$]. The imaginary part of $\sigma$ (Fig.\ \ref{Fig8}a) is very similar in both models, thus indicating that the frequency at which thermoplasmons are expected should be well described by Eq.\ (\ref{wj}). However, damping associated with electron-hole-pair transitions in the local-RPA model produces larger values of ${\rm Re}\{\sigma\}$ ({\it i.e.}, optical losses), which arise from the interband term in Eq.\ (\ref{local}).

Considering an array of 10\,nm ribbons (Fig.\ \ref{Fig8}e), we find the absorbance features predicted with the local-RPA conductivity for $E_F^0=0$ and finite $T$ (solid curves) to be broadened with respect to the absorbance for $T=0$ using the values of $E_F$ given by Eq.\ (\ref{EFT}) (dashed curves), but still leading to observable plasmon resonances. Notice that although the intrinsic damping is set to $\hbar\tau^{-1}=50\,$meV, the strong dispersion of $\sigma$ broadens the dashed-curve spectra in Fig.\ \ref{Fig8}e ({\it i.e.}, for finite $E_F$ and $T=0$) to 65, 65, and 74\,meV (from left to right), whereas thermal effects produce solid-curve spectra ({\it i.e.}, for finite $T$ and $E_F=0$) with widths of 212, 165, and 147\,meV. In a practical experiment, a finite distribution of ribbon widths in a self-assembled array \cite{HTT13,CRJ10,LTA13} would introduce further broadening. Alternatively, given the large expected levels of absorption, transient plasmons should be observable in individual structures. A more detailed account of the momentum dependence of $\sigma$ beyond local response theory could introduce some extra broadening, although the results of Fig.\ \ref{Fig3} indicate that this effect should be minor. Additionally, the electron gas temperature varies over a subpicosecond time scale, thus producing further broadening due to the change in the plasmon energy during its lifetime ({\it e.g.}, the lifetime is $\sim5\,$fs for the red solid spectrum of Fig.\ \ref{Fig8}e, which is a small interval compared with recently measured relaxation times \cite{BTM13}, and therefore, this should produce just a small broadening).

\section{Outlook and Perspectives}

Graphene has opened new perspectives in plasmonics research due to a combination of several appealing properties. From a practical viewpoint, its resistance to ambient conditions (particularly when encapsulated in between two dielectrics), its high degree of crystalinity, and its excellent electrical properties are well suited to the design of optoelectronic devices. Additionally, this material presents large optical nonlinearities \cite{HHM10,M11,NPA12}, as well as extraordinary electro-optical tunability \cite{CPB11}. In a more speculative front, the extreme confinement of graphene plasmons relative to the light wavelength gives rise to strong interaction with neighboring optical emitters, such as molecules and quantum dots \cite{paper176}, which has prompted several suggestions for the exploitation of this robust material to realize quantum optics phenomena in the solid-state environment of integrated gating devices \cite{paper184,HNG12,paper204,paper226}.

However, graphene faces important challenges that must be overcome before it can legitimately claim its privileged position among the zoo of plasmonic materials. An important challenge concerns the fabrication of patterned graphene structures with better control over shape and quality. While the number of methods that are becoming available to synthesize this atomically thin material is continuously growing \cite{NFC12}, a detailed tailoring of atomic edges will likely rely on bottom-up approaches, among which decoration of vicinal surfaces \cite{HTT13} and chemical self-assembly \cite{CRJ10,LTA13,BLZ12,LYG11} appear to be promising solutions. Alternatively, extended graphene can be inhomogeneously doped by patterning an underlying backgate \cite{VE11}, giving rise to confined and guided plasmons, as well as plasmon trapping at $p$-$n$ junction lines \cite{MSS10}.

In- and out-coupling to external light is another challenge. Progress in the former has been made through the realization that plasmon-assisted complete optical absorption is possible upon patterning monolayer graphene \cite{paper182}, followed by the recent experimental observation of high absorption \cite{paper212,paper230}. However, light emission from plasmon modes is intrinsically limited by their high degree of spatial confinement, which makes inelastic attenuation the dominant decay channel. As a possible solution, larger out-coupling to light could be achieved, with some limitations \cite{FK12}, through placing the carbon structure in an optical cavity in order to boost the density of optical states.

Graphene plasmons have been so far measured down to mid-IR wavelengths, including a spectacular full-octave range of electro-optical tunability in the mid-IR \cite{paper212}. Light modulation through graphene gating has also been observed down to vis-NIR frequencies in the optical plasmonic response of neighboring metal nanostructures \cite{ECN12,YKG13,LY13}, although the resulting degree of modulation is rather limited. This situation presents yet another challenge: the extension of octave-scale graphene confined-plasmon tunability to the vis-NIR spectral region, which could have massive impact on optical signal processing and telecommunications technologies. Some possible solutions to this problem have been put forward in the preceding sections.

Graphene is expected to exhibit a low level of inelastic optical losses compared with traditional plasmonic materials. We have discussed above the leading mechanisms that are thought to be responsible for such losses. However, an accurate experimental and theoretical determination of the ultimate intrinsic level of losses in high-quality graphene is still missing, in spite of some recent successes on both fronts \cite{JBS09,PVC13,paper212,BJS13}. This will obviously require further improvement of fabrication methods, alongside careful analyses of the relative importance of the noted mechanisms.

Advances in all these challenges may transform some of the widely advertised expectations raised by the plasmonics community into a fruitful reality. For example, in applications to optical sensing. Actually, IR plasmons cover the characteristic frequency range of molecular vibrations. Given the large enhancement of the optical field in their vicinities \cite{paper216}, graphene nanostructures could be used to reduce the concentration threshold for ultrasensitive molecular detection via infrared absorption spectroscopy. Inelastic light scattering enhanced by the graphene plasmon near-field is another possible strategy to improve optical sensitivity.

Quantum optics with graphene could rely on the use of solid-state two-level emitters such as NV centers in diamond, which could undergo quantum strong-coupling phenomena when resonantly coupled to graphene plasmons \cite{paper176,paper184}. Small graphene islands could actually act as two-level systems themselves, in which quantum nonlinearity produced by the combination of intrinsic graphene nonlinearity and strong plasmon field confinement has been predicted to create significant two-photon interactions \cite{paper226}.

In a different front, the detailed mechanisms responsible for photoelectric generation in graphene are still unclear, with both direct charge-carrier separation and the thermoelectric effect contributing to the light-induced electrical signal \cite{GSM11}. In this context, extrinsic metal plasmons have been used to increase the photoresponse of graphene \cite{LCL11,EBJ11} and demonstrate a nanoscale spectrophotometer \cite{FLW12}. The use of intrinsic graphene plasmons to this end could clearly allow us to spectrally resolve the incident light intensity by electrically tuning the plasmons in a graphene nanostructure, with some initial advances already made in this direction \cite{FLZ14}. This seems to be a realistic approach towards a nanoscale spectrophotometer, which could operate down to the mid-IR spectral range, and possibly also in the vis-NIR if graphene plasmons are successfully pushed towards higher frequencies. Incidentally, nanoscale tunable lighting devices could also be made of graphene based upon thermal emission, particularly in the IR, whereby the emission spectrum is proportional to the light absorbance (Kirchhoff's law), which is thus enhanced at the plasmon frequencies.

These ideas configure the exciting emerging field of graphene plasmonics. But perhaps the true impact of this activity lies in the realization that ultrathin materials can sustain tunable collective optical oscillations. The quest for such new materials has only started \cite{LBK13,SSS13}, with new exciting results already observed for plasmons in topological insulators \cite{DOL13}.

\appendix

\section{Quantum-Mechanical TB+RPA simulations}
\label{Ap1}

We follow a procedure described elsewhere \cite{paper183} to simulate nanostructured graphene by using its tight-binding (TB) electronic structure as input of the RPA susceptibility. This approach is computationally inexpensive and yields results in good agreement with state-of-the-art {\it ab initio} methods \cite{paper215}. The hopping parameter $t=2.8\,$eV is taken from a fit to both STM measurements \cite{PBV11} and first-principles simulations \cite{RMT02}. Notice however that the Fermi velocity \cite{W1947} $v^{\rm TB}_F=3ta/2\hbar$ extracted from $t$ and the C-C bond distance $a$ is $\sim10$\% lower than the electronic-band measured velocity \cite{BOS07} $v_F=10^6\,$m/s. We use this latter value for $v_F$ to relate $E_F=\hbar v_F\sqrt{\pi n}$ to the carrier density $n$, but we maintain the above value of $t$ in our TB+RPA simulations. The CNT band structures (Fig.\ \ref{Fig6}e) and $E_F(n)$ relations (Fig.\ \ref{Fig4}c) are obtained by counting TB states, so a factor $v_F/v_F^{\rm TB}$ is applied to the TB calculated $E_F$ in Fig.\ \ref{Fig4}c to compensate for this discrepancy. For simplicity, the hopping parameter is assumed to be the same for all carbon bonds, which should be a reasonable approximation when the edges are passivated with hydrogen atoms.

\section{Classical electromagnetic simulations}
\label{Ap2}

We follow a procedure sketched elsewhere \cite{paper181} to simulate the optical response of graphene ribbons using as input the frequency-dependent local conductivity $\sigma(\omega)$. More precisely, we describe the graphene through a square array of polarizable elements with in-plane polarizability given by $\alpha=1/[g/a^3-\ii\omega/(a^2\sigma)]$, where $a$ is the array period (small compared with the ribbon width) and $g=4.52$ results from the dipolar interaction summed over the lattice. With this choice of the polarizability, the reflection coefficients of an infinitely extended array coincide with those of homogeneous graphene in the electrostatic limit for both $s$ and $p$ polarizations, provided $a$ is sufficiently small. In practice, we obtain converged results for $a\sim1\,$nm. Using the $\exp(\ii k_\parallel z)$ spatial dependence of the fields along the ribbon direction $z$ for fixed parallel wave vector $k_\parallel$, we can carry out the lattice sum along $z$ and reduce the self-consistent system to just one row of polarizable elements across the ribbons. The results of Fig.\ \ref{Fig2} are converged using $\sim$100 elements and the computation of the entire figure takes only a few seconds using a personal computer. This method is in excellent agreement with numerical simulations using the boundary-element method for graphene ribbons described as thin slabs \cite{paper176}.

\section{Analytical model for the absorption cross-section of CNTs}
\label{Ap3}

We consider a single-wall CNT of small diameter $D$ compared with the light wavelength, illuminated under the conditions depicted in the upper inset of Fig.\ \ref{Fig6}a. We express the optical electric field $\Eb=-\nabla\phi$ in terms of the electrostatic potential, which is $\phi^{\rm in}=AR\cos\varphi$ inside the tube and $\phi^{\rm out}=-R\cos\varphi+(B/R)\cos\varphi$ outside, where $(R,\varphi)$ are polar coordinates in the plane normal to the tube. This is the most general solution of Poisson's equation for illumination with an external potential $-R\cos\varphi$, which corresponds to an incident unit electric field along $x$. Here, $A$ and $B$ are constants that are determined by the boundary conditions, namely, the continuity of the parallel electric field, $\partial_\varphi\phi^{\rm in}=\partial_\varphi\phi^{\rm out}$, and the jump in the normal electric field due to the surface current, $\partial_R\phi^{\rm in}-\partial_R\phi^{\rm out}=(16\pi\ii\sigma/\omega D^2)\partial_{\varphi\varphi}\phi^{\rm in}$. From here, we find the induced current to reduce to $A\sigma\sin\varphi\,\hat{\varphi}$, with $A=-1/(1+4\pi\ii\sigma/\omega D)$. Calculating the far field produced by this current with the help of the retarded Green function of the electromagnetic field, and using the optical theorem, we finally obtain the absorption cross-section of Eq.\ (\ref{sigmaCNT}).

\begin{figure}
\begin{center}
\includegraphics[width=70mm,angle=0,clip]{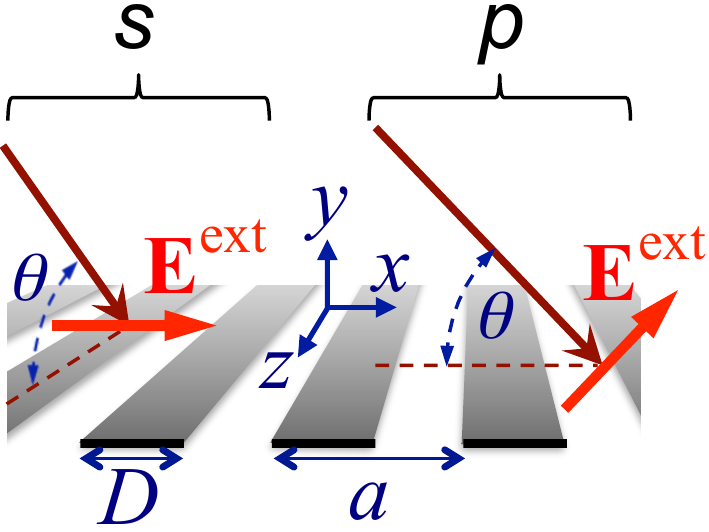}
\caption{Oblique incidence conditions here considered: external plane wave with wave vector in the $yz$ plane and electric field along $x$ ($s$ polarization, left); and incident wave vector and external field both in the $xy$ plane ($p$ polarization, right). Medium 1 (medium 2) is above (below) the ribbon array.} \label{Fig9}
\end{center}
\end{figure}

\section{Absorption by an array of ribbons under oblique incidence}
\label{Ap4}

We consider two special incidence conditions (see Fig.\ \ref{Fig9}): (1) incident light wave vector and surface-projected electric field both perpendicular to the ribbons long axis ($p$ polarization); and (2) incident electric field parallel to the surface and directed across the ribbons ($s$ polarization). Light is coming from medium 2 in both cases. In particular, Fig.\ \ref{Fig7}e is obtained for $s$ polarization. Describing the ribbons through their polarizability per unit length $\alpha_\omega/L$ and following the analytical methods that we introduced elsewhere \cite{paper090}, we find expressions for the resulting reflection and transmission coefficients similar to those already obtained for 2D periodic arrangements of graphene islands \cite{paper212}. More precisely,
\begin{align}
&r_s=r_s^0+\frac{iS_s(1+r_s^0)}{(\alpha_\omega/L)^{-1}-G_s},
&t_s=t_s^0+\frac{iS_st_s^0}{(\alpha_\omega/L)^{-1}-G_s},\nonumber\\
&r_p=r_p^0-\frac{iS_p(1-r_p^0)}{(\alpha_\omega/L)^{-1}-G_p},
&t_p=t_p^0+\frac{iS_pt_p^0}{(\alpha_\omega/L)^{-1}-G_p},\nonumber
\end{align}
for $s$ and $p$ polarizations, where
\begin{align}
&r_s^0=\frac{k_{\perp1}-k_{\perp2}}{k_{\perp1}+k_{\perp2}},
&t_s^0=\frac{2k_{\perp1}}{k_{\perp1}+k_{\perp2}},\nonumber\\
&r_p^0=\frac{\epsilon_2k_{\perp1}-\epsilon_1k_{\perp2}}{\epsilon_2k_{\perp1}+\epsilon_1k_{\perp2}},
&t_p^0=\frac{2\sqrt{\epsilon_1\epsilon_2}k_{\perp1}}{\epsilon_2k_{\perp1}+\epsilon_1k_{\perp2}}\nonumber
\end{align}
are the Fresnel coefficients of the graphene-free $\epsilon_1|\epsilon_2$ interface,
\begin{align}
&S_s=\frac{4\pi}{a}\frac{k^2}{k_{\perp1}+k_{\perp2}},
&G_s\approx\frac{g}{a^2}\frac{2}{\epsilon_1+\epsilon_2}+iS_s,\nonumber\\
&S_p=\frac{4\pi}{a}\frac{k_{\perp1}k_{\perp2}}
{\epsilon_2k_{\perp1}+\epsilon_1k_{\perp2}},
&G_p\approx\frac{g}{a^2}\frac{2}{\epsilon_1+\epsilon_2}+iS_p,\nonumber
\end{align}
$k_{\perp j}=(k^2\epsilon_j-k_\parallel^2+i0^+)^{1/2}$ (with ${\rm Im}\{k_{\perp j}\}>0$) and $k_\parallel=k\sin\theta$ are the perpendicular and parallel components of the wave vector in media $j=1,2$ for an angle of incidence $\theta$, $a$ is the lattice period, and $g=2\pi^2/3$ (see above). Obviously, we are neglecting diffracted beams because the spacing $a$ is assumed to be much smaller than the light wavelength. We should stress that $\alpha_\omega$ is the electrostatic polarizability. The above expressions are derived under the assumption that $\epsilon_j$ and $k_{\perp j}$ are real in both media $j=1,2$. They are also valid under total internal reflection conditions ({\it i.e.}, when $k_{\perp2}$ is imaginary), with $G_s$ and $G_p$ redefined as
\begin{align}
&G_s\approx\frac{g}{a^2}\frac{2}{\epsilon_1+\epsilon_2}+\frac{\ii S_s}{1-k_{\perp2}/k_{\perp1}},\nonumber\\
&G_p\approx\frac{g}{a^2}\frac{2}{\epsilon_1+\epsilon_2}+\frac{\ii S_p}{1-\epsilon_2k_{\perp1}/\epsilon_1k_{\perp2}}.\nonumber
\end{align}
These conditions are actually considered in Fig.\ \ref{Fig7}e.

\section*{}
\section*{Acknowledgement}

This work has been supported in part by the European Commission (Graphene Flagship CNECT-ICT-604391 and FP7-ICT-2013-613024-GRASP).

\bibliographystyle{apsrev}
\bibliography{../../bibtex/refs}

\end{document}